\documentclass[twocolumn]{aastex62}
\usepackage{amssymb}
\usepackage{amsmath}
\usepackage{amssymb}
\usepackage{empheq}
\usepackage{mathrsfs}
\usepackage{wasysym} 
\usepackage{fourier-orns}
\usepackage{adforn}

\usepackage{etoolbox}
\usepackage{lmodern}

\newcommand{\be}{\begin{align}}
\newcommand{\ee}{\end{align}}

\newcommand{\appropto}{\mathrel{\vcenter{
  \offinterlineskip\halign{\hfil$##$\cr
    \propto\cr\noalign{\kern2pt}\sim\cr\noalign{\kern-2pt}}}}}

\begin{document} 

\title{\bf Origins of Compact Mean-Motion Resonances: \\ \smallskip Evidence for Long-Range Migration and the Case of Kepler-36}

\author[0000-0002-7094-7908]{Konstantin Batygin}
\affiliation{Division of Geological and Planetary Sciences, California Institute of Technology, Pasadena, CA 91125}

\author[0000-0001-8476-7687]{Alessandro Morbidelli}
\affiliation{Laboratoire Lagrange, Universit\'e C\^ote d'Azur, Observatoire de la C\^ote d'Azur, CNRS, CS 34229, F-06304 Nice, France}
\affiliation{Coll\`ege de France, Universit\'e PSL, 75005 Paris, France}

\begin{abstract}
The observed census of resonant extrasolar planets spans a tantalizing display of orbital architectures, ranging from familiar 2:1 and 3:2 mean-motion commensurabilities to nearly co-orbital configurations characterized by period ratios close to unity. While mean-motion resonances are widely recognized as signposts of convergent disk-driven migration, the process through which the most compact systems are established remains puzzling, since resonance capture must repeatedly fail at a series of first-order commensurabilities before finally succeeding at a high resonant index. Motivated by this discrepancy, here we develop an analytic theory that fuses the stability-based resonance capture criterion with the conventional paradigm of active accretion disks and the standard model of type-I migration. Within this framework, we derive an expression for the stellocentric radius of resonance capture, $r_{\rm{c}}$, and show that it depends only on the product of the disk viscosity parameter, $\alpha$, and the opacity-contributing small-grain mass fraction, $f_\mu$. Applying this formalism to Kepler-36 -- the most compact known resonant system with a 7:6 period ratio -- we find that resonance locking could not have been established near the disk's inner edge. Instead, capture must have occurred at $r_{\rm{c}}\approx 1-4$\,AU, implying orbital decay of the planetary pair by approximately an order of magnitude. Viewed in this light, compact resonant architectures provide the clearest evidence for long-range migration among sub-Jovian planets. Moreover, the emerging picture is fully consistent with formation models in which super-Earths accrete within localized rings of planetesimals at orbital distances comparable to those that gave rise to the terrestrial planets of the Solar System.
\end{abstract}



\section{Introduction}
\label{sec:intro} 

Theoretical understanding of planetary origins traces a long and storied history, dating back at least to the Nebular Hypothesis \citep{Swedenborg,Kant,Laplace} and continuing through increasingly rigorous analytical treatments of the 20th century (see \citealt{1905ApJ....22..165M, 1943ZA.....22..319W, Jeffreys1949, 1967ARA&A...5..267T, Safronov1969, 1981PThPS..70...35H, 1982P&SS...30..755S, 1993ARA&A..31..129L} for reviews). Over the past three decades, however, a decisive observational shift -- driven by the discovery and characterization of extrasolar planets -- has unveiled the empirical outcome of planet formation throughout our Galaxy. Particularly, it is now broadly understood that short-period, multi-Earth-mass planets constitute the typical product of planetary conglomeration, orbiting approximately 50\% of FGK stars \citep{2011ApJ...742...38Y, 2013Sci...340..572H}. In light of this observational clarity, the classical question of how, precisely, this ubiquitous process of planet formation unfolds, has gained renewed prominence.


A number of important clues are offered by the aggregate population of known exoplanets. For instance, the increasingly well-populated mass-radius diagram of sub-Jovian planets shows that objects stripped of their hydrogen-helium atmospheres approximately follow a Murnaghan-type $R/R_{\oplus} \propto (M/M_{\oplus})^{1/4}$ mass-radius relationship \citep{2016ApJ...819..127Z, 2017AJ....154..109F}. This scaling points toward predominantly silicate-rich compositions, implying that the conglomeration of super-Earths likely occurs within regions interior to the water-ice sublimation line (typically presumed to lie at distances of about 3--5\,AU within their natal disks; see e.g., \citealt{2011ARA&A..49..195A, 2011ARA&A..49...67W, 2015A&A...575A..28B, 2024Icar..41716085Y}). Moreover, the observed inter-system uniformity -- often referred to as the ``peas-in-a-pod'' pattern \citep{2018AJ....155...48W, 2021ApJ...920L..34M} -- suggests a self-regulating accretion mechanism that yields planets of comparable mass within a given system. 

Stepping beyond the empirical patterns of masses and radii, multiple lines of evidence indicate that disk-driven migration \citep{1997Icar..126..261W} plays a fundamental role in shaping planetary architectures. For instance, the near-universal inner edges of planetary systems, seen as a break in the occurrence rate distribution at an orbital period of a few days \citep{2018AJ....155...89P}, coincide closely with theoretical predictions for the magnetospheric truncation radii of protoplanetary disks (\citealt{BatAdBeck23}; see also \citealt{2017ApJ...842...40L}). In turn, the associated sharp transitions in the surface density profiles create a local region where planet-disk interactions are expected to equilibrate \citep{Masset2006}. Likewise, the mere existence of mean-motion resonances, which require slow, convergent orbital evolution for their establishment \citep{Peale1976,HenrardLemaitre1983}, bears witness to the radial drift of planets through the nebular gas\footnote{It is worth noting that while the ``breaking the chains" early dynamical evolution scenario suggests that a substantial fraction of the exoplanetary census may have originated within resonant configurations \citep{Izidoro2017,2022AJ....163..201G}, the reach of migration required to produce most observed resonant systems remains essentially unconstrained, since resonance capture merely demands \textit{some} degree of orbital convergence \citep{2020MNRAS.495.4192C}, and exhibits virtually no dependence on the initial conditions.}. Nonetheless, the \textit{extent} of orbital migration -- whether characterized by modest radial shifts of order unity or orbital decay spanning orders of magnitude -- remains difficult to constrain from empirical grounds.

\medskip
\centerline{\adforn{21}}
\medskip

\begin{figure}
\includegraphics[width=\columnwidth]{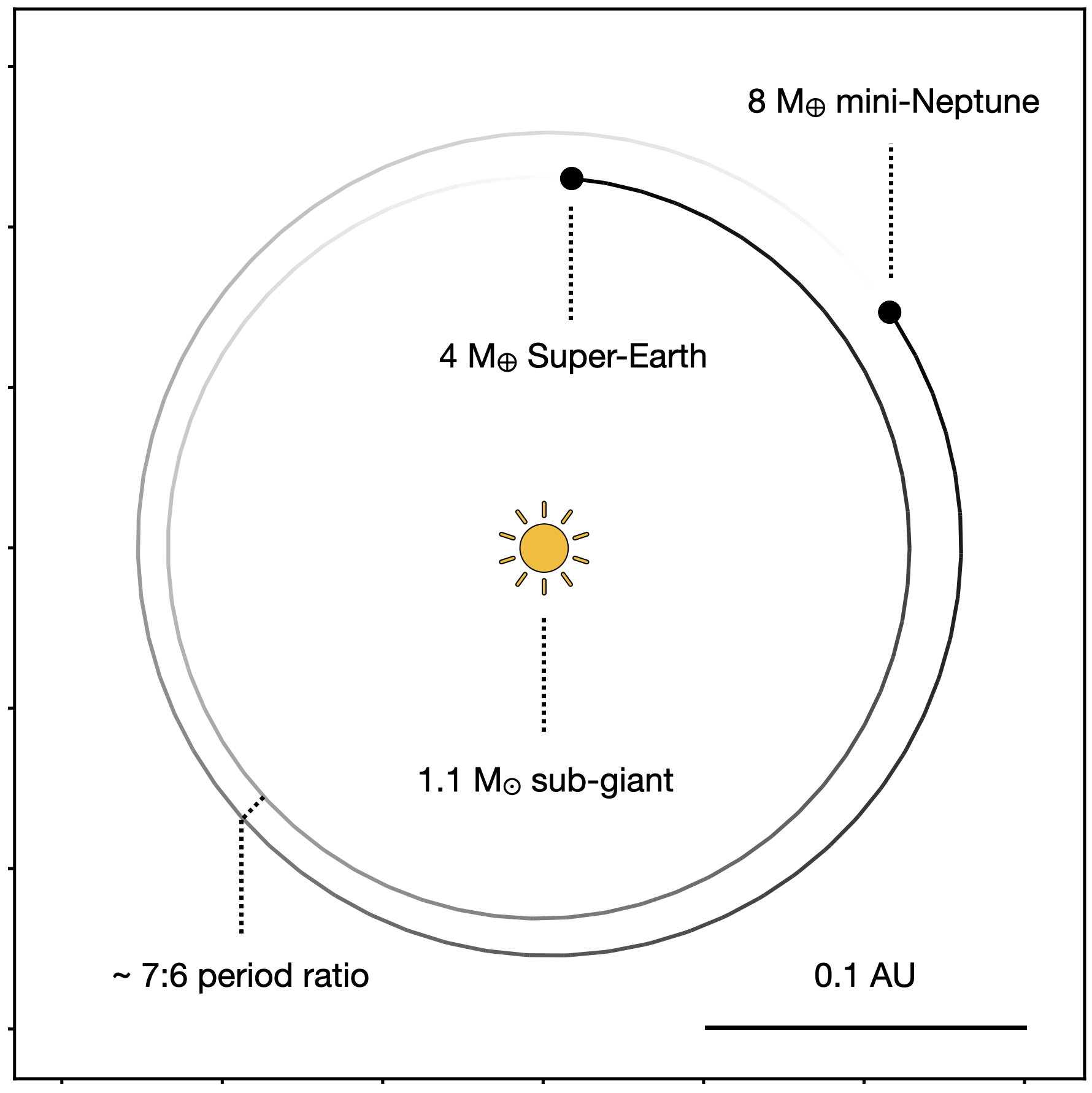}
\centering
\caption{Schematic view of the Kepler-36 planetary system. A $\sim4\,M_{\oplus}$ super-Earth (planet b) and a $\sim8\,M_{\oplus}$ mini-Neptune (planet c) orbit a $\sim1.1\,M_{\odot}$ subgiant host star at orbital periods of approximately 14 and 16 days. The planets’ semi-major axes differ by only $\sim 0.01\,$AU, placing them near the 7:6 mean-motion commensurability.}
\label{fig:K36}
\end{figure}

To begin addressing this central uncertainty, here we focus on a particularly rare class of systems: resonant planets locked into compact orbital states. The most striking example of such a configuration is presented by the Kepler-36 system. In terms of physical properties, the Kepler-36 planets -- a super-Earth of $m_1\sim4\,M_{\oplus}$ and a mini-Neptune of $m_2\sim8\,M_{\oplus}$ orbiting a $M = 1.07\,M_{\odot}$ star -- appear relatively unremarkable. With periods of approximately 14 and 16 days, however, the planetary orbits lie close to the 7:6 commensurability (Figure \ref{fig:K36}) and exhibit rapid dynamical chaos\footnote{Kepler-36 is not entirely unique in this regard. For instance, the Gliese 876 system also exhibits chaotic orbital evolution with a similarly short Lyapunov timescale of roughly 7 years \citep{2015AJ....149..167B}.}, which is driven by the overlap among the neighboring 29:34 high-order resonances, instilling the system with a Lyapunov timescale of a mere decade \citep{Deck2012}.

Dating back to its discovery and initial dynamical characterization \citep{2012Sci...337..556C}, Kepler-36’s unusually compact architecture has stood as an intriguing challenge for planet formation theory. Since then, numerous studies have attempted to unravel its origin. \citet{2013MNRAS.434.3018P} conducted hydrodynamical simulations of rapidly migrating planets and concluded that under specific assumptions, a laminar disk must exceed the mass of the minimum mass solar nebula by over an order of magnitude for the planets to bypass lower-index resonances (e.g., 2:1, 3:2) and end up stably trapped in the 7:6 resonance. Additionally, they found that turbulence could aid orbital compactification by stochastically disrupting lower-index resonances. \citet{2018MNRAS.479L..81R} carried out population-synthesis type $N$-body simulations of this scenario, suggesting that objects within sufficiently massive disks can indeed become locked into a 7:6 mean-motion resonance, while attributing the density contrast between the interior and exterior planets to their respective formation inside and outside of the nebular ice line.

\citet{Quillen2013} employed semi-analytic and numerical methods, similarly invoking turbulence to disrupt resonances, but highlighted the extreme fine-tuning required: turbulence must be sufficiently vigorous to break the 6:5 resonance but simultaneously gentle enough to preserve the 7:6 resonance\footnote{As shown in \citet{BatAd2017} from analytic grounds, the dependence of the turbulent disruption threshold on the resonant index is weak.}. Moreover, \citet{Quillen2013} noted that the turbulent sculpting scenario demands rapid disk dissipation almost immediately following resonant capture -- a condition that conflicts with the concurrent necessity of a relatively massive nebula. \citet{2021MNRAS.501.4255R} proposed a distinct inward-migration scenario accompanied by planetesimal collisions to explain Kepler-36’s compact architecture, and demonstrated that capture into the 7:6 commensurability is achievable in principle. However, the reported successes are intrinsically rare (at the percent level) and typically require eccentricity-damping efficiencies for the two planets that differ by large factors, in tension with the near-uniform scaling expected under conventional type-I prescriptions. Collectively, these studies represent important progress, but also underscore that a fully robust explanation for the dynamical origins of compact resonant architectures remains outstanding.


\medskip
\centerline{\adforn{21}}
\medskip


In this study, we leverage recent analytic advances to address the origin of compact resonant architectures. As we demonstrate below, capture into high-index resonances is regulated primarily by \textit{dissipative stability} in the presence of disk-driven eccentricity damping and convergent migration, a regime that admits closed-form criteria and is amenable to analytic treatment. This framing of the problem allows us to place concrete limits on the extent of migration experienced by the Kepler–36 planets. 

The remainder of this manuscript is organized as follows. In Section 2, we develop the relevant capture thresholds and validate them with a large suite of direct $N$–body simulations. In Section 3, we embed these criteria in an actively accreting disk by connecting to the standard type-I migration scalings, and obtain a closed expression for the capture radius $r_{\rm c}$. We then derive associated bounds on $\dot{M}$, $h/r$, $\Sigma$, and turbulent survival, consistent with formation interior to the ice line. Section 4 discusses the implications of our theory for planet formation and migration, addresses rarity and representativeness, and outlines observationally falsifiable predictions that can adjudicate among competing scenarios.



\bigskip
\bigskip

\section{Resonance Capture and Stability Under Dissipation}

The central question we aim to address in this section is: \textit{under what conditions does stable capture into the 7:6 resonance occur, subsequent to a series of failed encounters with lower-index (namely, 6:5, 5:4, etc.) commensurabilities?} To this end, we first delineate the general framework of disk-driven migration, and then enumerate the principal analytic criteria that determine whether resonance locking can occur. Building upon these foundations, we map the capture domains through direct numerical simulations and conclude by identifying the margin for the onset of overstability.

\subsection{Orbital Migration}

To proceed quantitatively, we begin by sketching out expressions for the disk-induced orbital decay timescales of the semi-major axis ($\tau_{a_j}$) and eccentricity ($\tau_{e_j}$). Standard equations for these timescales, widely employed in the literature, have been derived by \citet{2002ApJ...565.1257T, 2004ApJ...602..388T}, and can be written as:
\begin{align}
&\tau_{a_j} = \chi_{a} \left( \frac{m_j}{M} \frac{\Sigma\,a_j^{2}}{M} \left( \frac{h}{r} \right)^{-2} \Omega_j \right)^{-1}  \nonumber \\
&\tau_{e_j} = \frac{\chi_{e}}{\chi_{a}} \left( \frac{h}{r} \right)^{2} \tau_{a_j},
\label{tauataue}
\end{align}
where $j$ is the index of the planet, $\Sigma \propto r^{-s}$, is the disk's surface density, $a$ is the semi-major axis, $\Omega$ is the orbital frequency, $h/r$ is the disk aspect ratio, while $\chi_{e} = 1/0.78$ and $\chi_{a} = 1/(2.7+1.1\,s)$ are dimensionless constants, the latter of which depends on the power-law index of the surface density profile, with $s\sim\mathcal{O}(1)$. 


In a disk hosting two simultaneously migrating planets, it is useful to introduce the characteristic timescale of orbital convergence, together with the associated eccentricity-damping parameter:
\begin{align}
&\tau_{a} = \bigg( \frac{1}{\tau_{a_2}} - \frac{1}{\tau_{a_1}} \bigg)^{-1}  &\mathcal{K}_j = \frac{\tau_{a}}{\tau_{e_j}}.
\end{align}
The timescale $\tau_{a}$ can be rendered dimensionless by multiplying it by the orbital frequency of the inner planet, $\tau_{a}\,\Omega_1$ (note that for compact orbital configurations, $\tau_{a}\,\Omega_1\approx \tau_{a}\,\Omega_2$ and in some instances within the manuscript, we will simply drop the index for simplicity). Under the type-I migration prescription (equation~\ref{tauataue}), the parameter $\mathcal{K}_j$ depends primarily on the disk aspect ratio, scaling as $\mathcal{K}_j \propto (h/r)^{-2}$.

\subsection{Analytic Capture Criteria}

\paragraph{Adiabaticity \, } \ \ The process of resonant capture has been a subject of investigation for over half a century \citep{1965MNRAS.130..159G, 1966IAUS...25..197M, 1970MNRAS.148..325S}, and several well-established criteria have emerged from this extensive body of work. Foremost among them is the \textit{adiabaticity criterion}, which stipulates that resonance capture can only occur if the timescale over which disk-driven orbital migration transports planets across the width of the resonance is longer than the resonance libration period itself \citep{2001ApJ...547L..75F}. In the limit of compact orbits, this criterion reads \citep{Batygin2015}:
\begin{align}
\frac{1}{\tau_a\,\Omega} \lesssim  \frac{5\,\pi}{8}\bigg(\frac{5}{36\,k^5\,(k-1)^{2/3}} \bigg)^{1/3}\bigg(\frac{M}{m_1+m_2} \bigg)^{4/3},
\label{adiabaticcapture}
\end{align}
where $k$ is the resonance index ($k=7$ for the 7:6 MMR).

\paragraph{Eccentricity \, } \ \ Even if the adiabatic condition is satisfied, resonance capture can only be assured if the planetary orbits involved possess sufficiently low eccentricities (see e.g., \citealt{HenrardLemaitre1983, 1986sats.book..159P}). Given our assumption that resonance capture occurs within a protoplanetary disk -- an environment that naturally maintains nearly-circular planetary orbits -- this condition is effectively guaranteed, and therefore does not pose a practical limitation in our analysis.


\paragraph{Dissipative Stability \,}\ \ Beyond the classical criteria described above, an additional and somewhat more subtle limitation arises in the dissipative three-body problem. In resonance, convergent migration excites eccentricity while disk tides damp it, and long-lived trapping requires the existence of a fixed point in which these effects balance. As such, the resonant forcing must be sufficiently strong to offset eccentricity damping.

In general, this balance shifts the equilibrium away from the conservative resonant center. However, the resonant torque is bounded: the phase shift cannot grow without limit (in fact, it cannot exceed $\sim\pi/2$), and if eccentricity damping is too strong relative to the resonant forcing, no stationary resonant equilibrium exists. In this case, capture is impossible, and the system passes through the commensurability. The corresponding \textit{dissipative stability criterion} that delineates the existence of this stable resonant equilibrium was recently derived\footnote{See also the work of \citet{2023MNRAS.522..828H} for a related derivation in the circular restricted limit.} by \citet{2023ApJ...946L..11B}, and takes the following explicit form:

\begin{align}
\frac{1}{\tau_{a}} \lesssim \frac{32\,G\,k^{3}(m_{1}+m_{2})\,(m_{2}\tau_{e_1}+m_{1}\tau_{e_2})}{25\,M\,a_{2}^{3}},
\label{capture}
\end{align}
where $G$ is the universal gravitational constant. Rewriting this relationship in non-dimensional form yields:
\begin{align}
\frac{1}{\tau_{a}\,\Omega } \lesssim \frac{m_2}{M}\sqrt{ \frac{32\,k^{3}\,(1+\zeta)\,(\zeta\,\mathcal{K}_1+\mathcal{K}_2)}{25\,\mathcal{K}_1\,\mathcal{K}_2}},
\label{capturenondim}
\end{align}
where $\zeta = m_1/m_2$ is the planetary mass ratio.



\begin{figure*}
\includegraphics[width=0.75\textwidth]{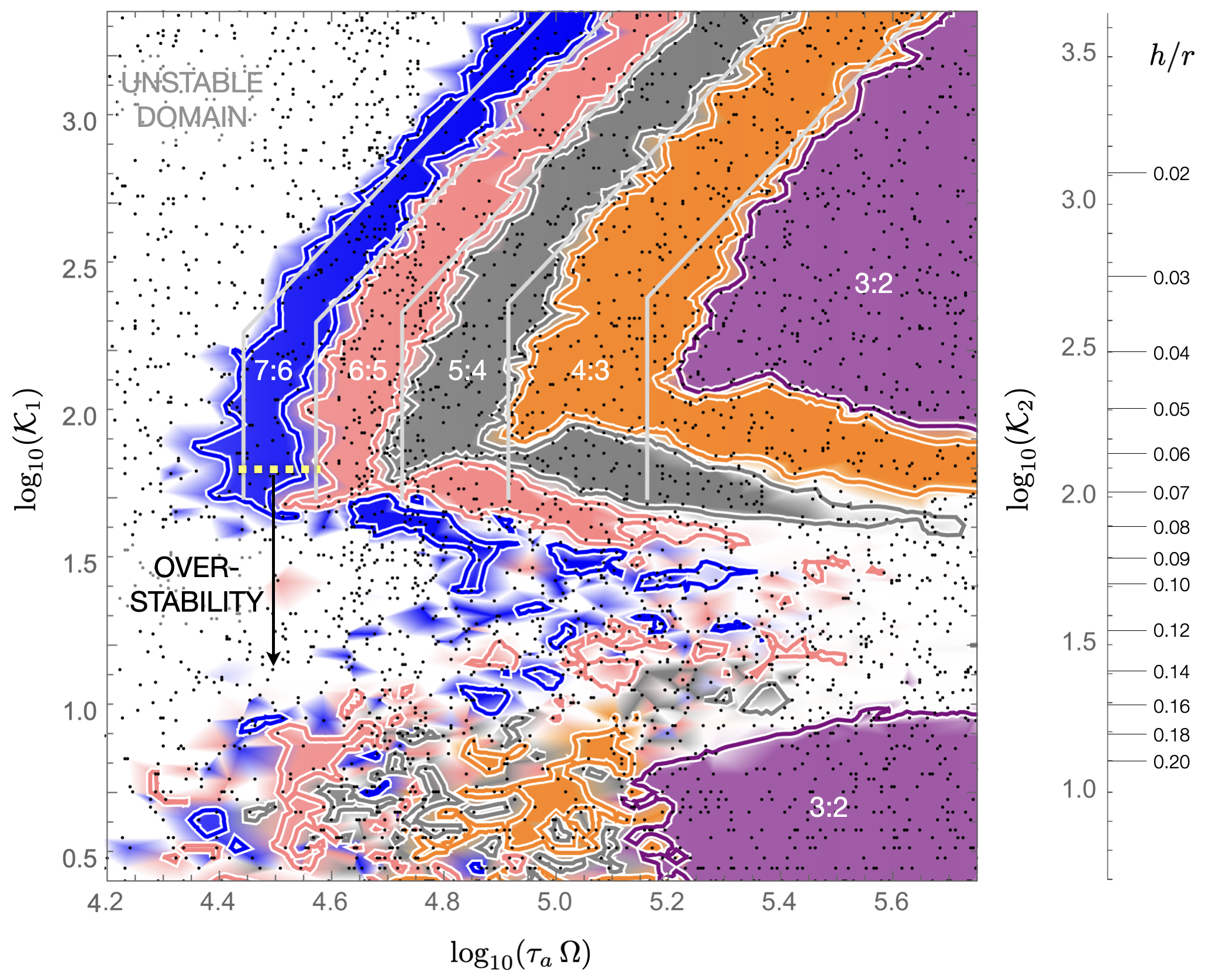}
\centering
\caption{Numerically computed resonance capture map. The $x$-axis shows the dimensionless product of the semi-major axis convergence timescale and the inner orbital frequency, while the $y$-axes show the ratios of eccentricity-to-semi-major-axis-damping times for the inner ($\mathcal{K}_{1} = (\tau_{a}/\tau_{e_1})$, left) and outer ($\mathcal{K}_{2} = (\tau_{a}/\tau_{e_2})$, right) planets. The masses of the bodies are fixed to the observed values of the Kepler-36 system. In the compact limit, this ratio directly reflects the disk aspect ratio $h/r$, also indicated on the right-hand scale (computed with $s=3/5$). Gray lines denote theoretical capture boundaries: vertical lines mark the adiabaticity threshold, while slanted lines show the dissipative stability criterion. Each point represents a distinct $N$-body calculation and the color bands correspond to regions of successful numerical capture into the labeled first-order resonance, while the white domain indicates unstable configurations. The overstability boundary for the 7:6 resonance is shown with a yellow dashed line.}
\label{fig:capture}
\end{figure*}

\subsection{Numerical Simulations}

To quantify the union of the above criteria from numerical grounds, we carried out a suite of direct $N$-body simulations, accounting for planet-disk interactions in a parameterized manner. In particular, we performed our integrations employing the Bulirsch-Stoer method \citep{1992nrfa.book.....P} and augmented gravitational dynamics with dissipative effects (mimicking migration and circularization) using the formulae of \citet{2000MNRAS.315..823P}. In these calculations, the planets were initialized on circular, planar orbits, with an initial period ratio of 2.5, and allowed to migrate convergently in presence of eccentricity (and inclination) damping.

For simplicity of numerical implementation, we renormalized the semi-major axes of the planets at every time-step, such that the inner orbit was always at radial distance of unity, and set the trial (initial) time-step to approximately 1/32nd of the inner orbital period. The integration accuracy parameter was set to $\hat{\epsilon}=10^{-13}$ for all runs. We remark that while this renormalization procedure amounts to an iterative change of units, this issue is fully circumvented by working in terms of dimensionless variables $\tau_a\,\Omega$ and $\mathcal{K}=\tau_a/\tau_e$.

In total, we carried out 3894 simulations with the dimensionless migration timescale $\tau_a\,\Omega$ spanning the $10^{4} - 10^{6}$ range, and $\mathcal{K}_1$ ranging from $10^{0.4}$ to $10^{3.4}$. The rate of eccentricity damping was assumed to be linearly proportional to the planetary mass (as entailed by equation \ref{tauataue}), fixing the value of $\mathcal{K}_2$ (for the outer planet) to $\mathcal{K}_1\times(m_2/m_1)= \mathcal{K}_1/\zeta$. Both the migration and circularization parameters were sampled in an approximately log-inform manner, and the duration\footnote{We have checked that shortening the duration to $3\,\tau_a$ does not change the results meaningfully.} of each run was set to $10\,\tau_a$.

The key results of our numerical experiments are shown in Figure (\ref{fig:capture}), where  the terminal period ratio attained by the planets is depicted on the $(\tau_a\,\Omega,\mathcal{K})$ plane. In particular, each point in the graph denotes an individual simulation and the interpolated heat-map marks the end-state commensurability\footnote{For definitiveness, we adopted equation \ref{reswidthchi} in the Appendix for the width of a resonance in terms of the planetary period-ratio.}. The transparency of each color corresponds to the fraction of simulations that end up in a given resonance, with the separating contours designating the 75th percentile. The domain with the white background represents cases where the integration results in an instability (quantified by a semi-major axis ratio attaining a value in excess of 5). 

A number of crucial features emerge on this diagram. In the upper half of the plot, resonance capture follows a banded pattern with progressively higher-index commensurabilities corresponding to decreasing semi-major axis convergence time. These boundaries, separating the various resonances, exhibit an approximately quadratic relationship between $\tau_a\,\Omega$ and $\mathcal{K}$ (i.e., $\mathcal{K}\appropto (\tau_a\,\Omega)^2$) for large $\mathcal{K}$, but their profiles become essentially vertical (corresponding to $\tau_a$ independent of $\mathcal{K}$) for more moderate levels of damping. Importantly, the last stable resonance in this sequence is the 7:6 MMR.


Both of these scaling relationships can be understood from analytic grounds. As discussed in \citet{2023ApJ...946L..11B}, between dissipative stability (equation \ref{capturenondim}) and adiabaticity (equation \ref{adiabaticcapture}), resonance capture is dictated by the more stringent of the two criteria. Following this line of reasoning, we plot contours of the more restrictive condition for $k=2,3,\dots,7$, as gray lines on Figure~\ref{fig:capture}. Indeed, these analytic boundaries satisfactorily track the numerical results, especially for $k\gg 1$.

\subsection{Overstability}

It is important to note that the aforementioned regime of stable resonance capture does not extend to arbitrarily low values of $\mathcal{K}$. Instead, for $\mathcal{K}_1 \lesssim 10^{1.7}$, resonance locking routinely gives way to the onset of dynamical instability\footnote{An exception ensues at very low values of $\mathcal{K}$ and large $\tau_a$, where capture into the 3:2 resonance once again becomes probable. However, this regime is likely irrelevant to protoplanetary disks.}. The threshold for this transition, too, can be understood from analytic grounds, and corresponds to the onset of overstability of resonant librations. This effect arises in systems where semi-major axis damping depends upon eccentricity, and under certain conditions can lead to a long-term growth of the libration amplitude. Originally developed in the test-particle limit of the circular restricted three-body problem by \citet{1995Icar..115...47G,Goldreich2014}, a more general criterion for the onset of overstability that accounts for non-zero masses and eccentricities of both planets was subsequently derived by \citet{2015ApJ...810..119D}; see also \citet{2015A&A...579A.128D, 2025ApJ...991...15B}.


In the limit of compact orbital configurations ($a_{1}/a_{2}\to 1$), this criterion depends explicitly on the resonance index $k$, the planet-to-star mass ratio, and the ratio of eccentricity damping to semi-major axis convergence timescales (as already discussed above, for type-I migration, this fraction translates to a dependence on the disk’s aspect ratio $h/r$). Rewriting the \citet{2015ApJ...810..119D} instability criterion in our notation, we obtain

\begin{align}
\left[\frac{\zeta \,\big(k(1+\zeta)-2\zeta\big)}{(1-\zeta)} \,\frac{\chi_{a}}{\chi_{e}}\right]^{3/2} <\, \frac{15}{32}\,\frac{1-\zeta^{2}}{M/m_{2}} \left(\frac{h}{r}\right)^{3}.
\label{eqn:overstability}
\end{align}
For the masses of the Kepler-36 planets, this criterion dictates that $k=7$ overstable librations are expected to ensue for $h/r \gtrsim 0.063$. Though marginally higher than typically quoted aspect ratios of inner regions of mature protoplanetary disks ($h/r\lesssim 0.05$), this threshold establishes a dynamical limit on the possibility of long-term resonance maintenance\footnote{This limit is not absolutely sharp: even overstable systems may avoid the onset of scattering by entering into a limit-cycle state; however, the available parameter space for such behavior is restricted.}. 

Taken together, the results of our numerical experiments presented in Figure~\ref{fig:capture}, can be interpreted as a natural consequence of the intersection among the three principal resonance-capture criteria: dissipative stability, adiabaticity, and overstability. In the regime of strong damping ($\mathcal{K}\gg1$), dissipative stability governs the character of resonant encounters, delineating the boundary between capture and passage. As the degree of dissipation decreases, this regime gradually yields to a narrow domain where resonance capture becomes limited by adiabaticity, below which the onset of overstable librations suppresses long-term locking altogether. Notably, the transition from dissipation-dominated to adiabatic behavior occurs near values of $\mathcal{K}$ corresponding to the upper end of plausible protoplanetary disk parameters. Consequently, under disk conditions likely to be realized in practice, the dynamics of resonance capture are expected to be regulated primarily by dissipative stability, and for the remainder of this work, we focus on this physically motivated regime.



\section{Resonant Capture Radius and Disk Constraints}

With the relevant criterion for establishing high-index resonances outlined and numerically validated, let us now examine how it maps onto the properties of an actively accreting protoplanetary disk by explicitly connecting the analytic framework to the standard model of type-I migration. In particular, we begin by merging the resonance-capture stability criterion with the governing relations for disk structure to express the capture radius in terms of fundamental nebular parameters. We then consider additional constraints arising from the disk’s thermal properties -- most notably the requirement that the inner planet form interior to the water-ice line. As we show below, this condition imposes a lower bound on the mass accretion rate, which in turn yields corresponding limits on the disk’s aspect ratio and surface density, culminating in an assessment of the circumstances under which turbulent fluctuations can disrupt resonance maintenance. To set the stage, we start by recalling the explicit functional form of the dissipative stability criterion (\ref{capture}).

Upon substitution of equations (\ref{tauataue}) into expression (\ref{capture}), the inequality can be re-written as:
\begin{align}
r^{4} \lesssim \frac{32\,k^{3} M^{2} \left( 1 + \zeta + \zeta^{2} + \zeta^{3} \right) \chi_{a} \chi_{e}}{25 \left( 1 - \zeta \right) \zeta\, \Sigma^{2}} \left( \frac{h}{r} \right)^{6}
\label{capturemigration}
\end{align}
Notably, we have ignored the variation in $\Sigma$, $\Omega$, and $(h/r)$ between the two orbits. Given their extreme proximity to one-another, this approximation does not entail a substantial drawback for high-index commensurabilities (but introduces inaccuracies for lower values of $k$). 

In practice, $\Sigma$, $r$, and $h/r$ are not strictly independent parameters but mutually constrained descriptors of a protoplanetary disk's state, and their values reflect the coupled thermodynamic and transport properties of the nebula. As a result, substituting numerical estimates in isolation would obscure the physics at play. Instead, it is more illuminating to first reformulate the capture condition in terms of the global structure of protoplanetary disks, where these quantities are connected by well-established scalings. 

\subsection{Disk Structure}

Generally speaking, quantification of disk structure remains an active area of investigation, with significant ongoing debate regarding the relative roles of turbulent viscosity \citep{1974MNRAS.168..603L,1973A&A....24..337S} vs. magnetic effects \citep{1982MNRAS.199..883B, 2022ApJ...930..167Z} in facilitating angular momentum transport and determining disk evolution. Although these roles undoubtedly vary in both space and time \citep{2019SAAS...45....1A, 2023A&A...677A.136M}, most contemporary models agree on a broad, qualitative picture: viscous processes dominate angular momentum transport in the inner regions of the disk, particularly at early evolutionary stages when disks are comparatively massive. In contrast, outer disk regions tend to adopt a more passive, wind-dominated profile.

As one concrete illustration of this dichotomy, the self-similar disk solution of \citet{2022MNRAS.512.2290T}, which provides a unified treatment of turbulent and magnetic effects, effectively reduces to the classical viscous (actively accreting) disk solution at orbital radii interior to a characteristic transition radius $r\lesssim r_{\rm{t}} \sim 10 - 15\,\text{AU}$ \citep{2024Icar..41716085Y}. Given that the composition of the inner planet Kepler-36b is unequivocally silicate-rich, our analysis naturally focuses on regions interior to the water-ice sublimation line, where the active viscous disk approximation is expected to hold true. Accordingly, we adopt this well-studied framework throughout the remainder of our study.


The surface density is related to the mass accretion rate $\dot{M}$ via the well-known relation (e.g., \citealt{2011ARA&A..49..195A}):
\begin{align}
\Sigma = \frac{\dot{M}}{2\,\pi\,r\,v_r} = \frac{\dot{M}}{3\,\pi\,\nu} = \frac{\dot{M}}{3\pi\,\alpha \,h^2\,\Omega},
\label{sigmamdot}
\end{align}
where $v_r$ is the radial gas velocity and we have used the \citet{1973A&A....24..337S} $\alpha$ parameterization for the viscosity, $\nu$. Physically, $\alpha$ represents a dimensionless measure of the efficiency of angular momentum transport, and is often of order (within factors of unity) the square of the turbulent Mach number. In the inner disk, this transport can be supplied by a variety of (magneto-)hydrodynamic processes whose relative importance may vary with stellocentric distance and time, including the magnetorotational instability \citep{1991ApJ...376..214B}, the zombie vortex instability \citep{2015ApJ...808...87M,2016MNRAS.462.4549L}, convective overstability \citep{2003ApJ...582..869K,2014ApJ...788...21K}, and the vertical shear instability (\citealt{2013MNRAS.435.2610N}; see also \citealt{2019PASP..131g2001L} for a review). This assortment of processes is expected to yield a broad range of effective $\alpha$, spanning $\alpha \sim 10^{-4}$ -- $10^{-2}$. Here we adopt this interval as astrophysically plausible, while noting that values toward the lower end are likely more typical in the inner disk at temperatures below the thermal ionization threshold for MRI activation (conservatively, $T \lesssim 10^3\,{\rm K}$; \citealt{2015ApJ...811..156D}; see also \citealt{2025PSJ.....6....2Y}).

The relevant expression for the scale height, $h$, depends upon whether the disk is optically thick or thin. The former limit applies when the optical depth, $\bar{\tau}$, substantially exceeds unity:
\begin{align}
\bar{\tau} = \frac{1}{2}\,\Sigma \,f_\mu \,\kappa \gg 1.
\label{opticallythick}
\end{align}
In the above expression, $\kappa$ is the Rosseland-mean dust opacity and $f_\mu$ denotes the dust-to-gas ratio of the opacity-contributing grain population\footnote{Rosseland-mean opacity is typically controlled by grain sizes comparable to the peak of the local thermal emission, i.e., broadly micron to tens-of-micron sized particles.}. Strictly speaking, $\kappa$ depends on wavelength (and hence temperature and disk location) as well as on the grain size distribution \citep[e.g.,][]{2016A&A...586A.103W}. In the temperature range relevant to the inner disk (corresponding to $T\sim 200-10^3\,{\rm K}$), the opacity variations predicted by commonly used prescriptions are comparatively modest \citep{1994ApJ...427..987B}. For simplicity, we follow \citet{2015A&A...575A..28B} and adopt a representative constant value $\kappa \approx 30\,{\rm m^2\,kg^{-1}}$, with the understanding that residual radial and temporal changes in the effective opacity can be absorbed into the local value of the product $\kappa\,f_\mu$\footnote{While the canonical dust-to-gas ratio inferred from the interstellar medium is $\sim 1\%$, grain growth and planetesimal formation reduce the mass budget of small, opacity-contributing grains to values below this estimate. Nevertheless, the inner disk is not expected to become devoid of such grains: inward drift resupplies solids from larger radii, while collisional fragmentation replenishes the small-grain population that sets the Rosseland-mean opacity \citep{DominikDullemond2008,Birnstiel2010,Birnstiel2011,2016A&A...586A.103W}. Consequently, even if most of the solid mass resides in larger aggregates, the inner few AU are generically expected to remain optically thick.}. To this end, $f_\mu$ is expected to vary with both radius and time, and detailed models suggest that, like $\alpha$, $f_\mu$ plausibly spans $\sim 10^{-4}$ -- $10^{-2}$ in the inner disk \citep[e.g.,][]{2019ApJ...874...26S,2020A&A...637A...5E}.


It is worth noting that some studies fold an assumed small-grain abundance $f_\mu\simeq 10^{-2}$ into an ``effective'' opacity and therefore quote values of $\kappa$ that are smaller by approximately two orders of magnitude; here we instead factor $f_\mu$ explicitly, since it can deviate substantially from $10^{-2}$ during disk evolution. For the reference Minimum Mass Solar Nebula (MMSN)-like surface density of $\Sigma_0\approx2000\,\mathrm{g\,cm^{-2}}$, condition (\ref{opticallythick}) is satisfied, provided $f_\mu \gtrsim 3\times10^{-5}$.

Under this optically thick approximation, the mid-plane disk temperature has the form (see e.g., \citealt{2022A&A...666A..19B}):
\begin{align}
T = \left( \frac{3\,f_{\mu}\,\kappa\,\bar{\mu}\,\Omega^{3}\,\dot{M}^{2}}{64\,\pi^{2}\,k_{\rm{b}}\,\sigma_{\mathrm{sb}}\,\alpha} \right)^{1/5},
 \label{eqn:temperature}
\end{align}
where $\bar{\mu}\approx2.4\,m_p$ is the mean-molecular weight, $k_{\rm{b}}$ is the Boltzmann constant, and $\sigma_{\mathrm{sb}}$ is the Stefan-Boltzmann constant. In principle, stellar irradiation also contributes to the disk temperature profile at some level. However, for the regime relevant to our analysis, which is largely limited to radii interior to the ice line, the mid-plane energy budget is expected to be dominated by viscous dissipation for accretion rates characteristic\footnote{Although it is possible to posit sufficiently low accretion rates for irradiation to contribute non-negligibly to the thermal structure, such late-stage disks would also be characterized by substantially reduced surface densities. This would in turn imply correspondingly long type-I migration timescales, rendering the process of traversing a long sequence of low-index commensurabilities prior to capture into the 7:6 resonance unlikely.} of Class-II disks.

In this limit, the disk aspect ratio, $h/r=c_{\rm{s}}/v_{\rm{k}}\propto\sqrt{T}$, becomes:
\begin{align}
\frac{h}{r} = \left( \frac{3\,f_{\mu}\,k_{\rm{b}}^{4}\,\dot{M}^{2}\,\kappa}{64\,\pi^{2}\,\alpha\,\bar{\mu}^{4}\,\sigma_{\mathrm{sb}}} \sqrt{\frac{r}{G^{7} M^{7}}} \right)^{1/10}.
\label{aspectratio}
\end{align}
With the dependence of the aspect ratio on the radius specified, and assuming $\dot{M}$ independent of $r$ (steady state accretion), equation (\ref{sigmamdot}) simplifies to a $\Sigma\propto r^{-s}$ power law profile with an index $s=3/5$ (and we adopt this value for evaluation of $\chi_a$ for self-consistency).

\subsection{Resonance Capture Radius}

\begin{figure*}
\includegraphics[width=0.65\textwidth]{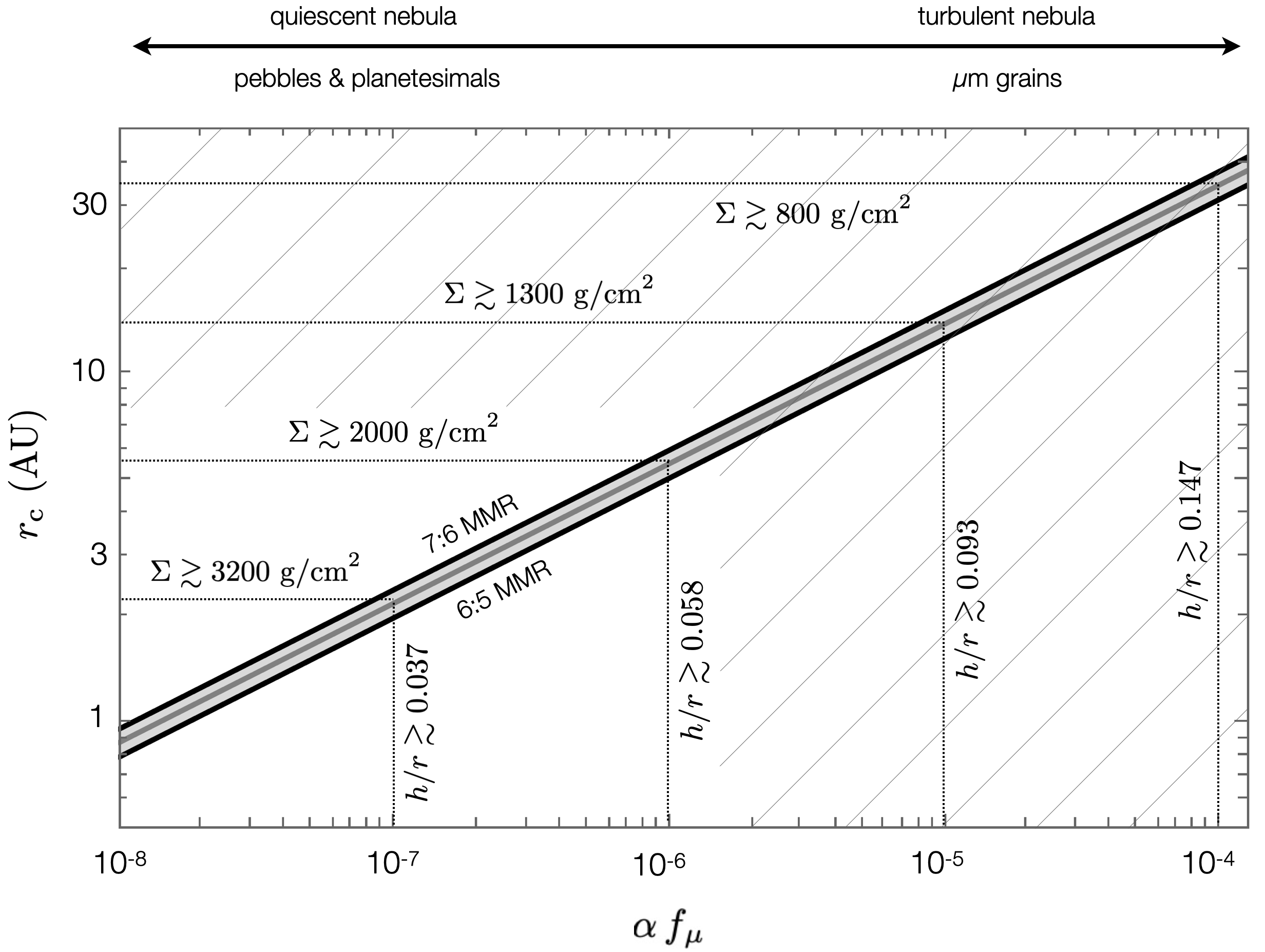}
\centering
\caption{The orbital radius for resonance capture, $r_{\rm{c}}$, as a function of the product $\alpha \, f_\mu$, where $\alpha$ is the disk viscosity parameter and $f_\mu$ is the dust-to-gas ratio of opacity-contributing small grains. The two black curves denote dissipative stability-based capture criteria for the 7:6 and 6:5 resonances, bracketing the admissible parameter range. Qualitatively, the left-hand-side of the diagram (with $r_{\rm{c}}\sim1-4\,$AU) corresponds to a parameter regime of a mature, relatively quiescent disk, where the small-grain dust-to-gas ratio has been dramatically reduced due to particle coagulation and planetesimal formation. The right-hand-side of the diagram denotes the opposite regime of a young, highly turbulent disk, where the majority of the solid budget of the nebula is in the form of small grains. The quantities annotating the dotted lines denote the minimal surface density, $\Sigma$ (horizontal labels) and aspect ratio, $h/r$ (vertical labels) -- both evaluated at $r_{\rm{c}}$ -- necessary for capture to occur interior to the water-ice sublimation line, so as to match the rocky composition of the inner planet. Regions of parameter space away from the smaller end of $r_{\rm c}$ and $\alpha f_\mu$ are disfavored: in this regime (marked as a hatched region), the implied surface densities and disk aspect ratios rise to unphysical values, and the corresponding resonant configurations are expected to be dynamically unstable due to overstability and turbulent disruption.}
\label{fig:rc}
\end{figure*}


Looking back to inequality (\ref{capturemigration}), recall that the surface density and the aspect ratio appear in the expression at inverse square and sixths powers, respectively. With their individual functional forms delineated in equations (\ref{sigmamdot}) and (\ref{aspectratio}), we find that within the term:
\begin{align}
    \frac{1}{\Sigma^{2}} \left( \frac{h}{r} \right)^{6} = \frac{27\,k_{\rm{b}}^{4}\,r^{3/2}\,\kappa\, \alpha\,f_{\mu}}{64\,(G M)^{5/2}\,\bar{\mu}^{4}\,\sigma_{\mathrm{sb}}},
\end{align}
the dependence on the accretion rate fully cancels out, leaving the product of viscosity parameter $\alpha$ and small-grain dust-to-gas ratio $f_\mu$ as the composite independent variable. Substituting this form back into inequality (\ref{capturemigration}) and rearranging once again for $r$, we obtain a criterion for resonance capture:
\begin{align}
r \lesssim r_{\mathrm{c}} = \Bigg(&\frac{32\,k^{3} M^{2} \left( 1 + \zeta + \zeta^{2} + \zeta^{3} \right) \chi_{a} \chi_{e}}{25 \left( 1 - \zeta \right) \zeta} \nonumber \\
&\times \frac{27\,k_{\rm{b}}^{4}\,\kappa}{64\,(G M)^{5/2}\,\bar{\mu}^{4}\,\sigma_{\mathrm{sb}}} \cdot \alpha\,f_{\mu}\Bigg)^{2/5}.
\end{align}
Adopting the ranges for $\alpha$ and $f_\mu$ discussed in the previous subsection (and noting that for Kepler-36, the admissible solution lies between $k=6$ and $k=7$ contours), in Figure (\ref{fig:rc}) we show the capture radius as a function of $\alpha\, f_\mu$, which we treat as an effective local quantity evaluated at $r_{\mathrm{c}}$.


The lower bound on $r_{\mathrm{c}}$ shown in Figure (\ref{fig:rc}) is of considerable interest. Specifically, below the line labelled 6:5, capture would occur into a resonance with index below $k<7$, implying that capture in the 7:6 resonance ($k=7$) can only occur for $r_{\mathrm{c}}\gtrsim 1\,\mathrm{AU}$. In turn, this indicates that the Kepler-36 planetary pair did not form locally and did not become captured into the compact resonance configuration it occupies today at the inner edge of the protoplanetary disk. Rather, the 7:6 resonance must have been established at a larger stellocentric distance, after which the system underwent inward migration while trapped in resonance and ultimately stalled at either the radius corresponding to the onset of the magneto-rotational instability or the disk's truncation radius, carved by the stellar magnetosphere \citep{2021A&A...648A..69A}. 

Physically, this limit on $r_{\rm{c}}$ follows directly from how disk torques scale with orbital radius. The underlying reason for why resonance capture cannot plausibly occur at the present-day orbital radii of the Kepler-36 planets is that the convergent migration rate, when normalized by the orbital frequency, \textit{declines} toward smaller stellocentric distances. That is, as $r$ diminishes, the quantity $\tau_{a}\,\Omega$ grows, meaning that successful bypass of resonances like 2:1 or 3:2 would require unrealistically short eccentricity-damping timescales (see Figure \ref{fig:capture}). In turn, the unusually low timescale ratio $\tau_e/\tau_a$ would correspond to a range of disk aspect ratios far smaller than what is astrophysically reasonable. 

This issue is further exacerbated by the effects of the inner edge of the disk itself, where the enhancement of corotation torques further suppresses the rate of convergent migration. Because the 7:6 configuration is nearly co-orbital, both planets would lie close to the equilibrium zone of planet--disk interactions, reducing their relative drift by a large factor compared with the power-law scalings embodied in equations (\ref{tauataue}). Accordingly, capture into so compact a resonance could only have been established at significantly larger stellocentric distances, and the requirement for long-range inward migration is in fact even more stringent than what the above analytic estimate suggests. Therefore, capture at high resonant index constitutes a direct observational proxy for long-range migration.



\bigskip
\bigskip

\subsection{Lower Bound on $\dot{M}$}

\begin{figure}
\includegraphics[width=\columnwidth]{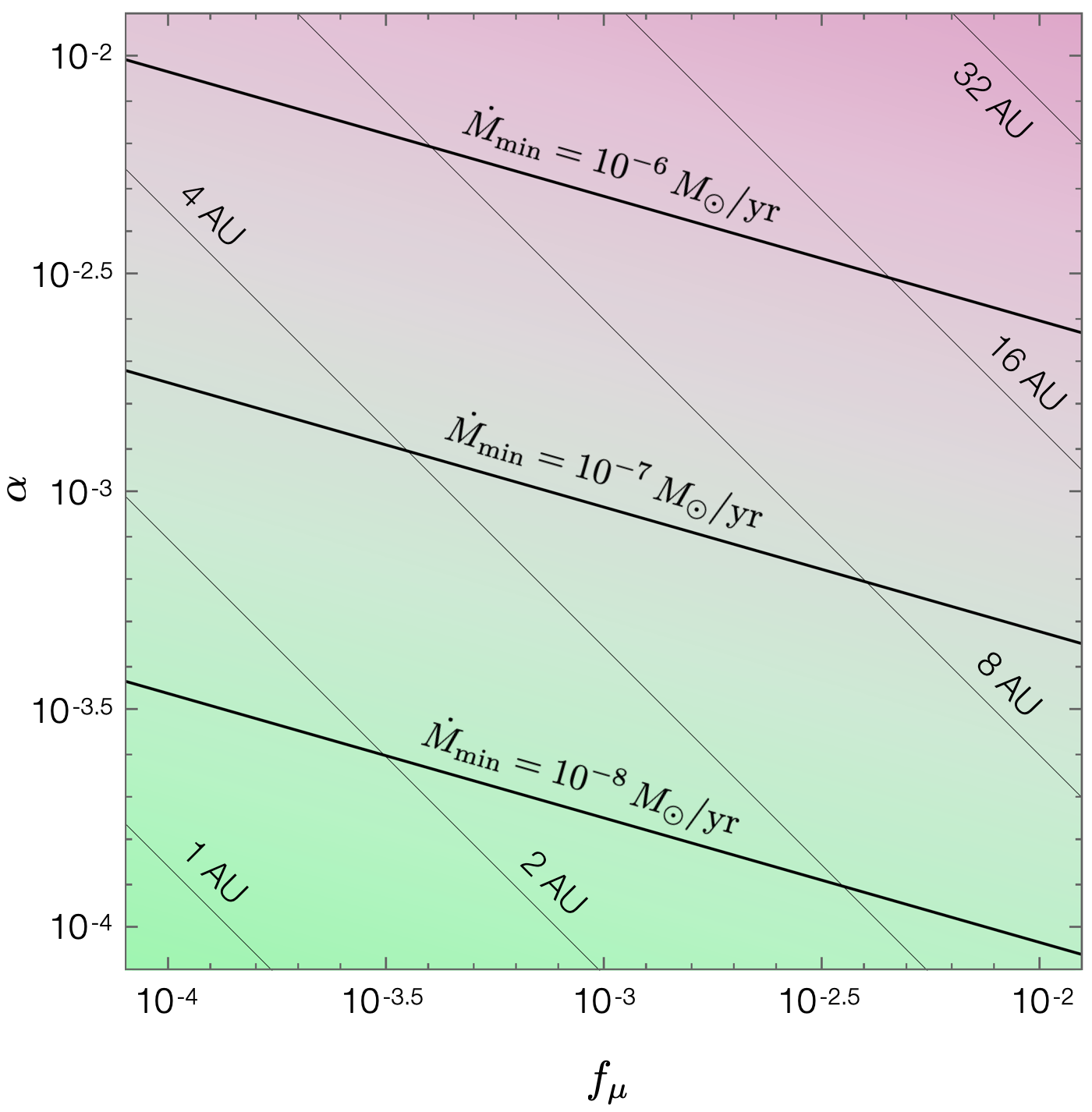}
\centering
\caption{Minimal disk accretion rate consistent with resonance capture within the nebular ice line. The thin lines show the resonance capture radius $r_{\rm{c}}$, which depends on the product $\alpha\,f_\mu$. The thick lines show the minimal requisite value of $\dot{M}$ such that, at the corresponding value of $r_{\rm{c}}$, the disk mid-plane temperature exceeds the water-ice sublimation temperature, $T_{\rm{ice}}$. Typical accretion rates in class-II disks are approximately bounded from above by $\dot{M} \lesssim 10^{-8} M_\odot/$yr, suggesting capture radii that correspond to the lower end of the admissible range (i.e., $1-2\,$AU, possibly up to $4\,$AU).}
\label{fig:Mdot}
\end{figure}


While the capture radius itself depends solely upon $\alpha$ and $f_{\mu}$, it is instructive to further explore the corresponding implications for the disk mass accretion rate, $\dot{M}$. Specifically, we begin by considering a fundamental limit imposed by disk thermodynamics. As previously noted, the rocky composition of Kepler-36b mandates that this planet formed interior to the water-ice sublimation boundary. Under typical protoplanetary disk conditions, this requirement translates into a constraint on the mid-plane disk temperature at distance of formation (given by equation \ref{eqn:temperature}), demanding that $T > T_{\mathrm{ice}}\approx170\,\mathrm{K}$. Given that inequality (\ref{capturemigration}) explicitly provides the orbital radius (and thus orbital frequency $\Omega$) as a function of $\alpha$ and $f_{\mu}$, this temperature threshold can be recast as a lower limit on the disk mass accretion rate, $\dot{M}_{\mathrm{min}}$:
\begin{align}
&\dot{M}_{\rm{min}} = \sqrt{\frac{64\,k_{\rm{b}}\,\pi^{2}\,T_{\mathrm{ice}}^{5}\,\sigma_{\mathrm{sb}}}{3\,\kappa\,\bar{\mu}\,\Omega^{3}}\, \frac{\alpha}{f_{\mu}}\, \,\bigg( \frac{r_{\mathrm{c}}^3}{G\,M} \bigg)^{3/2}}, 
\end{align}
where we have taken advantage of the narrowness of the admissible capture interval between the $k=6$ and $k=7$ resonances and set $k=13/2$ as a representative value for definiteness (the corresponding curve is shown on Figure \ref{fig:rc} with a thin gray line).

We illustrate this lower bound, along with corresponding level curves of the capture radius $r_{\mathrm{c}}$, in Figure (\ref{fig:Mdot}). In more specific terms, the three contours of constant $\dot{M}_{\mathrm{min}}$ depicted on the figure, imply that within the parameter space lying \textit{below} each thick curve, any mass accretion rate equal to or exceeding the labeled value is permissible. Correlating this constraint with the capture radius, it is noteworthy that the lower limit on $\dot{M}$ decreases with $r_{\mathrm{c}}$. Stated differently, for any fixed ratio $\alpha/f_{\mu}$, a decrease in the product $(\alpha f_{\mu})$ -- and consequently a reduction in $r_{\mathrm{c}}$ -- directly translates to a progressively lower $\dot{M}_{\mathrm{min}}$.

Generally, the mass accretion rates of disks diminish rapidly with age, with very high values of $\dot{M}$ confined to early ($t \lesssim 1\,\mathrm{Myr}$) evolutionary stages. Empirically derived $\dot{M}$-age relationships \citep{1998ApJ...495..385H, 2017MNRAS.468.1631R} imply that typical Class-II planet-forming disks are characterized by $\dot{M}\sim10^{-8}\,M_\odot\,\mathrm{yr}^{-1}$, corresponding to capture radii $r_{\mathrm{c}}\lesssim2-4\,\mathrm{AU}$. Although the precise epoch marking the onset of planet formation remains uncertain, this result suggests that large capture radii are disfavored.


\subsection{Aspect Ratio}

\begin{figure*}
\includegraphics[width=\textwidth]{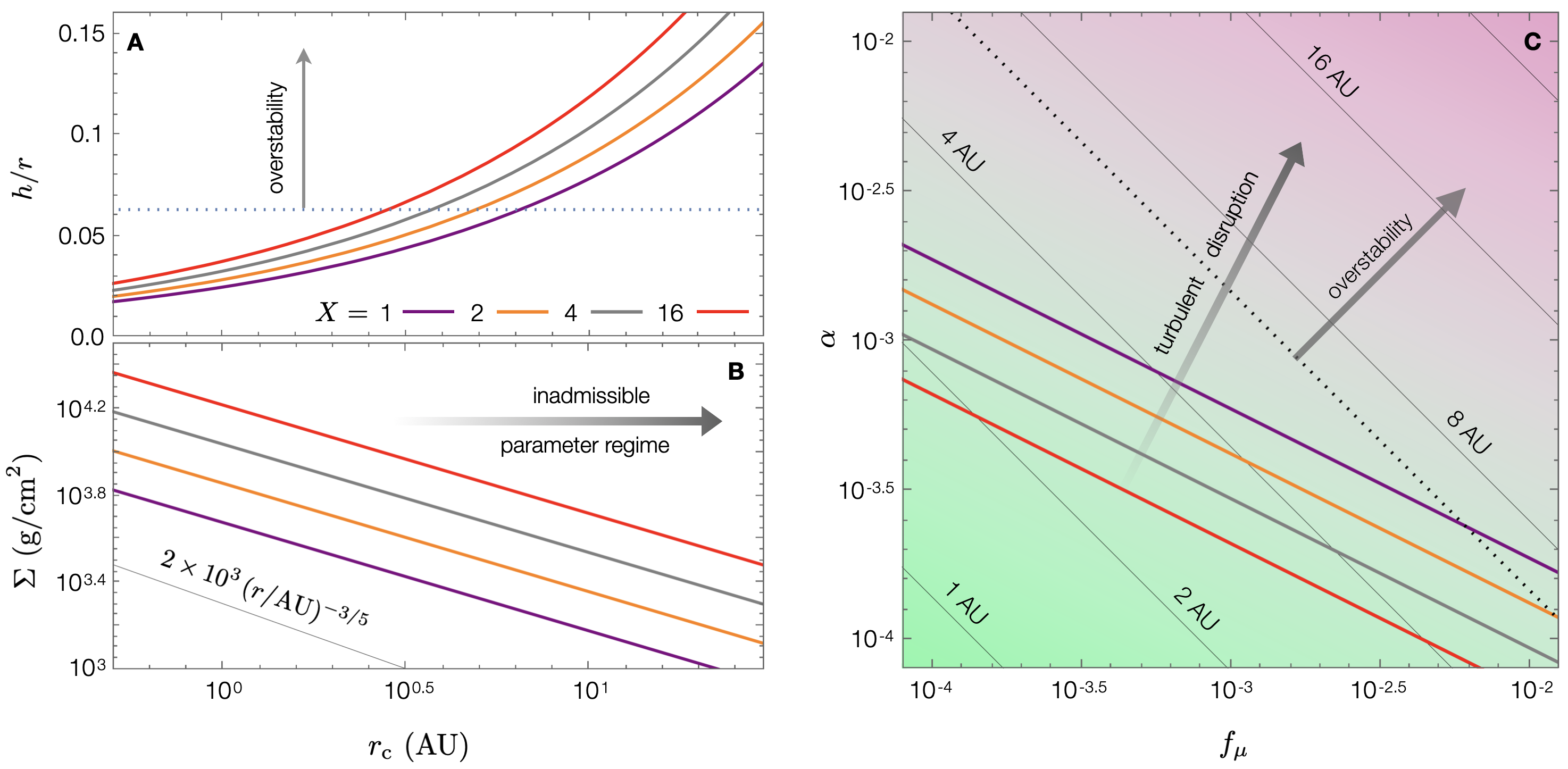}
\centering
\caption{Constraints on disk structure from resonance capture. Panel A shows the disk aspect ratio $h/r$ as a function of capture radius $r_{\rm{c}}$ for a range of the normalized accretion parameter $X=\dot{M}/\dot{M}_{\rm{min}}$, with the overstability threshold (equation \ref{eqn:overstability}) marked by a dotted line. Panel B depicts the corresponding surface density $\Sigma$, with a reference MMSN-like active disk profile plotted for comparison. Panel C shows an analytic stability map informed by turbulent disruption, projected onto the $\alpha - f_\mu$ plane. Thick lines denote limits beyond which turbulent disruption prohibits resonance maintenance for various values of $X$, while the dotted line denotes the onset of overstability. Together, these constraints favor resonance capture at $r_{\rm{c}} \lesssim 4\,$AU under moderately massive disk conditions.}
\label{fig:disk}
\end{figure*}

An equivalent calculation can be readily carried out for the disk aspect ratio $(h/r)_{\mathrm{min}}$. Employing the same ice-line temperature constraint as above, we obtain:
\begin{align}
\frac{h}{r} &= \Bigg( \frac{27\,f_{\mu}\,k^{3}\,X\,\alpha \left( 1 + \zeta + \zeta^{2} + \zeta^{3} \right) \kappa\,\chi_{a}\,\chi_{e}}{50\,M^{3}\,(1 - \zeta)\,\zeta\,\sigma_{\mathrm{sb}}}  \Bigg)^{1/5}  \nonumber \\
&\times \frac{\sqrt{T_{\mathrm{ice}}}}{G} \left( \frac{k_{\rm{b}}}{\bar{\mu}} \right)^{13/10} = X^{1/5}\,\sqrt{\frac{k_{\rm{b}}\,T_{\rm{ice}}}{\bar{\mu}}\frac{r_{\rm{c}}}{G\,M}},
\label{horX}
\end{align}
where we have introduced $X = \dot{M}/\dot{M}_{\rm{min}}\geqslant1$ as a measure of accretion for convenience.


This expression is shown as a function of $r_{\mathrm{c}}$ in Figure (\ref{fig:disk} panel A) for a range of scaled mass accretion rates and a number of representative lower-bound values are also labeled on Figure (\ref{fig:rc}). To this end, we note that although $h/r$ is not entirely independent of the disk accretion rate (in contrast to $r_{\mathrm{c}}$), the aspect ratio scales weakly, as $(h/r)\propto\dot{M}^{1/5}$. Conventional protoplanetary disk models (e.g., \citealt{2015A&A...575A..28B, 2024Icar..41716085Y} and references therein) typically find that for orbital radii within $\lesssim10\,\mathrm{AU}$, the aspect ratio is bounded from above by $(h/r)\lesssim0.05$. Consequently, the depicted constraint on $(h/r)$ further supports the scenario where resonance capture ensues at relatively smaller orbital radii, specifically $r_{\mathrm{c}}\lesssim4\,\mathrm{AU}$.

\subsection{Surface Density}

With constraints on both $\dot{M}$ and $(h/r)$ in hand, we can now employ equation (\ref{sigmamdot}) to derive an additional limit on the disk surface density:
\begin{align}
\Sigma &= \Bigg( \frac{2^{7} k^{9} k_{\rm{b}}^{7} X^{6} \left( 1 + \zeta + \zeta^{2} + \zeta^{3} \right)^{3} \sigma_{\mathrm{sb}}^{2} \chi_{a}^{3} \chi_{e}^{3}}
{15^{6} f_{\mu}^{2} M^{4} \alpha^{2} (1 - \zeta)^{3} \zeta^{3} \kappa^{2} \bar{\mu}^{7}} 
\Bigg)^{1/10} \nonumber \\
&\times \frac{4\,T_{\mathrm{ice}}^{3/2}}{G} = \frac{8\,X^{3/5}}{3} \sqrt{\frac{\bar{\mu}\,\sigma_{\rm{sb}}\,T_{\rm{ice}}^3}{3\,\alpha\,f_{\mu}\,k_{\rm{b}}\,\kappa}} \bigg(\frac{r_{\rm{c}}^3}{G\,M} \bigg)^{1/4}
\label{SigmaX}
\end{align}

Similarly to the aspect ratio, we depict a range of lower-bound values of $\Sigma$ in Figure (\ref{fig:rc}), and the surface density is shown as a function of $r_{\rm{c}}$ in Figure (\ref{fig:disk} panel B) for a sequence of normalized accretion rates, $X$. In addition, a reference power-law profile of an actively accreting disk, normalized to the MMSN value $\Sigma_0 \approx 2000\,\mathrm{g\,cm^{-2}}$ at 1\,AU, is plotted as a thin black line on the Figure. Our lower-limit estimate of the surface density at $r \sim 1$\,AU is larger than the MMSN value by a factor of $\sim2$. This suggests that $X$ is unlikely to exceed unity by more than a modest factor, since doing so would push the disk surface density well above canonical expectations.




\subsection{Turbulent Disruption}
A final bound we examine concerns the stability of the 7:6 resonance in the face of turbulent fluctuations within the protoplanetary disk's gravitational potential. \citet{BatAd2017} analyzed the dynamics of resonance capture within a turbulent nebula using a perturbative approach, and derived an analytic criterion for resonance stability as a solution to a stochastic differential equation. In this formulation, turbulent excitation is modeled as a drift-free Wiener process, while convergent migration acts as a damping term, resulting in bounded stochastic evolution (which can then be compared with the resonance width). Crucially, the resulting criterion explicitly depends upon $\alpha$, $\Sigma$, and $(h/r)$ -- quantities for which we have derived meaningful limits above.

In the Appendix, we present a modified derivation of this stability criterion, explicitly incorporating the migratory evolution of the resonant planet pair. Upon combining this criterion with the above-derived expressions for disk parameters, the resulting condition for stability can be succinctly expressed as:
\begin{align}
&\Bigg(\bigg(\frac{X\,f_{\mu}\,\kappa}{5\,\sigma_{\rm{sb}}} \bigg)^2 \bigg( \frac{k_{\rm{b}}}{\bar{\mu}} \bigg)^{13} \bigg( \frac{T_{\rm{ice}}\,\chi_a}{G^2} \bigg)^{5} \frac{(1+\zeta+\zeta^2+\zeta^3)^3}{(1-\zeta)^5\,\zeta^3}  \Bigg)^{1/4} \nonumber \\
&\times \frac{3\,\alpha}{25}\Bigg(\frac{2^9\,\chi_e^9\,k_{\rm{b}}^{23}}{m_{2}^{14}\,M^4\,(1+\zeta)^8} \Bigg)^{1/12} \lesssim 1.
\label{turbulentX}
\end{align}
Contours of this dimensionless stability quantity are illustrated in Figure (\ref{fig:disk} panel C) as thick multi-colored lines. As with the preceding constraints, this turbulent disruption criterion likewise favors capture radii at relatively modest stellocentric distances, specifically $r_{\mathrm{c}}\lesssim5\,\mathrm{AU}$. 

\medskip
\centerline{\adforn{21}}
\medskip


Collectively, the suite of constraints examined in this section suggests that the compact resonant architecture of Kepler-36 most naturally originates from formation and resonant capture at orbital radii exterior to $\sim1\,$AU but interior to $\sim2-4\,\mathrm{AU}$, followed by inward, disk-driven migration toward the stellar magnetospheric truncation radius. While the establishment of such a configuration requires a particular combination of surface density and aspect ratio, these conditions are not exotic. Rather, they fall squarely within the range expected for actively accreting protoplanetary nebulae — albeit toward the more massive end of canonical disk models.

\section{Discussion}

Investigations of resonant planetary systems have, by now, a long and productive history. Early studies of the Gliese 876 system \citep{2002ApJ...567..596L} first demonstrated clearly that eccentricity damping, operating in concert with convergent migration, plays a fundamental role in shaping observed resonant configurations. Subsequent investigations into multi-planet systems such as HD\,45364 \citep{2010A&A...510A...4R}, Kepler-223 \citep{2016Natur.533..509M, 2021A&A...656A.115H} and HD\,110067 \citep{2023Natur.623..932L} further illustrated how disk-driven orbital convergence and the degree of turbulence within the disk can directly sculpt resonant planetary chains. Other recent examinations of short-period resonant systems, such as Kepler-60, Kepler-80 \citep{2016MNRAS.455L.104G,2021AJ....162..114M, 2018MNRAS.477.1414C}, as well as the newly discovered resonant TOI-1136 system \citep{2023AJ....165...33D}, have expanded this perspective, highlighting the complex interplay between the rate of planet–disk interactions, resonant capture, and long-term dynamical evolution (to this end, see also \citealt{2024AJ....168..239D}). Along similar lines, the widely studied TRAPPIST-1 system has provided an illustrative case in which regression of the disk’s inner edge has played a significant role in the observed resonance chain \citep{2022MNRAS.511.3814H, 2024NatAs...8.1408P}. Finally, a select number of resonant systems (e.g., K2-19, Kepler-221), have been examined with an eye towards illuminating the role played by eccentricity and obliquity tides in affecting the observed physical and orbital architecture \citep{2020AJ....159....2P, 2020ApJ...897....7M, 2020MNRAS.496.3101P, 2021AJ....162...16G}.

Despite the breadth of insight gleaned from these studies, a fundamental question has remained unresolved: do planets form close to the disk’s inner edge, experiencing only limited orbital decay, or are they instead driven inward from considerably larger distances by extensive gas-driven migration? To address this question, in this work, we have considered the origins of the nearly-co-orbital architectures of compact resonant systems, adopting Kepler-36 as a case study. 

\subsection{Key Results}

To quantify the problem of high-index resonance capture, we began by outlining a series of analytic criteria, and validated them through a large suite of direct $N$-body simulations that incorporated convergent migration and eccentricity damping. The resulting capture map (Figure \ref{fig:capture}) reveals a structured sequence of stable commensurabilities, in which progressively higher-index resonances appear at shorter semi-major axis convergence timescales. The boundaries separating the various resonant domains are well reproduced by the analytic adiabaticity and stability criteria, confirming that the dissipative stability limit of \citet{2023ApJ...946L..11B} constitutes the pertinent condition for capture across most of the relevant parameter space (see also \citealt{2023A&A...669A..44K}). Importantly, these calculations also demonstrate that the 7:6 mean-motion resonance marks the last attainable stable commensurability for Kepler-36.

Building upon this framework, we combined the stability-based criterion with standard models of accretion disks and type-I migration to derive an explicit expression for the stellocentric radius at which resonance locking is expected to occur. Remarkably, this relationship collapses to a single-parameter sequence involving only the product of the disk $\alpha$-viscosity and the small-grain dust-to-gas fraction, $f_{\mu}$ (Figure \ref{fig:rc}). Application of this result to the Kepler-36 system indicates that the 7:6 resonance must have been established exterior to $\sim1$\,AU but interior to the nebular ice line, i.e., within $\sim4$\,AU.


While the full narrative of planetary growth, migration, and eventual stabilization at the disk’s inner edge is undoubtedly relevant to the Kepler-36 system, our analysis has focused more narrowly on the conditions for resonance capture. The reason for this focus is straightforward: among the various stages of the system’s evolution, the establishment of its compact resonant architecture is arguably the most puzzling. By contrast, post-capture evolution -- specifically how orbital migration slows and stabilizes at the cavity edge, etc. -- has been examined extensively \citep{Masset2006, 2018CeMDA.130...54P, 2021A&A...648A..69A}. We further note that after the dispersal of the nebula, additional processes such as tidal dissipation \citep{2012ApJ...756L..11L, 2013AJ....145....1B, 2019NatAs...3..424M} and anisotropic mass loss \citep{2015MNRAS.452.1743T} can further sculpt planetary architectures. While none of these mechanisms could have plausibly driven the system \textit{into} the 7:6 resonance from an initially larger period ratio, the latter mechanism has been previously suggested as one potential means to account for the mild observed \textit{departure} of the planets from exact 7:6 commensurability, once the disk had evaporated. 

\subsection{Implications for Planet Formation}

The results obtained herein have direct implications for the understanding of the principal mode of planetary accretion itself. Broadly speaking, contemporary theories of super-Earth formation fall into three categories, that differ considerably in the extent of migration they require. One class of theories envisions \textit{in-situ} assembly near the present locations of close-in planets \citep{2012ApJ...751..158H, 2013MNRAS.431.3444C, 2014ApJ...797...95L}, potentially invoking disk pressure maxima (such as those at MRI dead-zone boundaries) as traps for solids and embryos \citep{2014ApJ...780...53C, 2019A&A...630A.147F}. At the opposite extreme, pebble accretion and distributed-planetesimal models \citep{2019A&A...627A..83L, 2021A&A...656A..70E} permit accretion across extended regions of the disk, with growth efficiency markedly enhanced beyond the ice line \citep{2022A&A...666A..19B, 2024Icar..41716085Y}. Intermediate between these two regimes lies an emerging framework in which super-Earths are expected to originate within discrete rings of solids located near $\sim1$ AU \citep{2023NatAs...7..330B}. In this scenario -- recently expanded through sophisticated disk modeling by \citet{2024ApJ...972..181O} -- the concurrent action of particle growth and aerodynamic drift leads to the pile-up of solids in narrow annuli, seeding rapid accretion and producing the initial conditions for compact multi-planet systems.\footnote{See also \citet{2020ApJ...894..143B} and \citet{2022NatAs...6...72M} for related models of the formation of giant-planet satellites and the emergence of the NC-CC dichotomy among solar-system meteorites.}

Within this spectrum of possibilities, the order-of-magnitude migration range inferred for the Kepler-36 planets points most directly toward the ring-based formation picture. In this interpretation, the two planets originate within the same localized annulus of solids -- a scenario that naturally yields near-contemporaneous growth and therefore provides the initial conditions required for rapid convergent migration and resonant locking. Indeed, the characteristic migration timescale of the inner planet and the convergent migration timescale of the pair are both short ($\tau_{a_1} \sim \tau_a \sim 2-3 \times10^4\,\mathrm{yr}$ for our fiducial parameters), suggesting that reproducing the observed architecture likely requires the two planets to reach their migration-dominated phase within a timescale comparable to $\tau_a$. Such tight timing is a natural outcome of ring-driven accretion models, in which once a phase of rapid growth ensues, a number of multi-$\mathrm{M}_\oplus$ objects emerge within a narrow radial annulus on timescales of $\sim \mathrm{few}\times 10^4$ -- $10^5$ years \citep{2023NatAs...7..330B}.

After formation and capture, the pair migrates inward together before stalling near the disk’s inner cavity, preserving the compact resonant architecture. The striking density contrast between Kepler-36b and Kepler-36c can then be understood without invoking formation on opposite sides of the ice line: despite similar formation locations, the more massive planet can retain a significant H/He envelope while the less massive one is efficiently stripped by photoevaporation \citep{2013ApJ...776....2L}.  Moreover, this view naturally connects to terrestrial-planet formation within the Solar System, where analogous local concentrations of solids govern the assembly of rocky planets \citep{2022NatAs...6...72M,2025AJ....170..180N}.

\subsection{On Rarity and Representativeness}

Given that orbital configurations akin to that of the Kepler-36 planets are intrinsically rare, it may be tempting to dismiss the broader relevance of this system and suppose that its formation pathway was likewise exceptional. We find this view unlikely for several reasons. 

First, the physical properties of the Kepler-36 planets are entirely typical of the sub-Jovian population. Both bodies occupy the same region of the mass--radius diagram as the majority of super-Earths and mini-Neptunes. If the mechanism that produced them were truly exotic, it would be a remarkable coincidence for it to yield planets that are otherwise indistinguishable from the general census. 

Second, within the context of our theoretical framework, high-index resonances are \textit{expected} to be rare, even if the underlying mode of planet formation is common. Several stringent conditions must be satisfied simultaneously for capture into such compact commensurabilities to occur. (i) The accretion of both planets must proceed nearly contemporaneously, so that resonance locking can take place before the inner planet migrates too far inward. (ii) The mass ratio between the two planets must also fall within a specific range: the outer planet must exceed the inner planet’s mass by a significant margin to ensure rapid convergent migration, yet not so greatly that the system is immediately rendered over-stable (recall equation \ref{eqn:overstability}) or transitions into the slower, type-II migration regime. (iii) Perhaps most restrictively, the parameter space permitting stable capture into the 7:6 mean-motion resonance (shown in Figures \ref{fig:capture} and \ref{fig:rc}) is intrinsically narrow\footnote{Because this admissible domain is restricted, it is admittedly difficult to rule out that some compact resonant systems could potentially arise through genuinely low-probability,  stochastic channels \citep[e.g.,][]{2021MNRAS.501.4255R}. The distinguishing feature of our mechanism is that it is deterministic: when disk parameters fall in the appropriate regime, capture into a high-index commensurability is an expected outcome and yields a direct, testable link between compact resonances and the disk conditions at the time of assembly.}.

Beyond these arguments, post-nebular evolution further limits the survival of such configurations. Once the gaseous disk dissipates, resonant systems are prone to dynamical instability, with compact resonances being especially fragile. This is consistent with the ``breaking-the-chains'' scenario in which resonant chains assembled in the disk commonly destabilize after disk dispersal, washing out resonant period ratios in the mature population \citep{Izidoro2017,2022AJ....163..201G}. Observationally, resonant occurrence appears to decline with age \citep{2024AJ....168..239D}, suggesting that resonance survival is intrinsically uncommon even if resonance capture is frequent. Taken together, these considerations imply that the rarity of systems like Kepler-36 does not point to an uncommon formation pathway, but rather to the delicate nature of compact resonances themselves. The system’s architecture therefore provides a veritable window into a ubiquitous process, even if only a few examples survive to be observed.

\bigskip
\bigskip

\subsection{Concluding Remarks}

A distinguishing feature of any explanatory model relative to a descriptive one is the capacity to predict. In this vein, the framework developed herein offers a clear falsifiable criterion for when high-index resonant pairs should emerge. Among the various physical dependencies identified above, the planet–planet mass ratio provides the most direct observational lever: the theory requires that the planet-planet mass ratio, $\zeta$, be substantially smaller than unity. In this regime, convergent migration naturally operates at large stellocentric distances, enabling establishment of compact commensurabilities under disk conditions that are within canonical expectations. Conversely, for $\zeta \gtrsim 1$, migration is predicted to be divergent within the disk, and the only pathway toward nearly co-orbital architectures would involve assembly directly at the disk’s inner edge. In practice, this latter pathway demands unphysical disk parameters, and so compact resonances in such systems are not expected to arise.

As an illustrative test that is already implicitly present within the current exoplanet census, we may consider the Kepler-50 system. Although it is not trapped in a 7:6 resonance, the planets of this system lie close to the 6:5 commensurability. Existing mass determinations \citep{2024ApJS..270....8W} suggest that the planet-to-planet mass ratio in Kepler-50 is nearly identical to that of Kepler-36, and thus fully consistent with our theoretical expectations. However, current uncertainties in the measured masses remain large. The discovery of a $k\gg1$ resonant system characterized by $\zeta \gtrsim 1$ would stand in direct conflict with the framework presented here, demanding either revision of the theory or the existence of unusual disk structures capable of producing locally rapid convergent migration. Future discoveries of compact resonant chains will serve as the decisive test of the theoretical picture presented in this work.

\acknowledgments We are thankful to Max Goldberg, Ian Brunton, Teng Ee Yap, and Andrew Howard for insightful discussions. KB is grateful to the David and Lucile Packard Foundation, the Caltech Center for Comparative Planetary Evolution (3CPE), the National Science Foundation (grant number: AST 2408867) and NASA (Emerging Worlds grant number: 80NSSC26K0395) for their generous support. We thank the anonymous referee for a thorough and insightful report, which improved the quality of the manuscript.

\appendix

\section*{Turbulent Disruption of Resonances for a Migrating Planet Pair}


To leading order, the orbital effects of fully developed $\alpha$-turbulence on a resonant planetary pair can be modeled as a stochastic diffusion of the semi-major axis ratio away from its nominal commensurability value, with a diffusion coefficient given by \citep{2008ApJ...683.1117A, 2013ApJ...771...43O}:
\begin{align}
\mathcal{D}\approx\frac{\alpha}{2}\bigg(\frac{\Sigma\,a^2}{M} \bigg)^2\,\Omega.
\end{align}
In contrast, convergent migration acts as an effective restoring process, damping deviations of the semi-major axis ratio on a characteristic timescale $\tau_a$. The interplay between these two mechanisms -- stochastic diffusion driven by turbulence and deterministic damping driven by disk torques -- ultimately governs the resonant outcome: if turbulent fluctuations dominate, the resonance is disrupted; if migration-induced damping prevails, resonant locking is maintained and the pair remains phase-protected.

\citet{BatAd2017} examined the simplified case where both $\mathcal{D}$ and $\tau_a$ were held constant -- a setting appropriate for resonance capture near the inner edge of the disk. In the present context, however, where migration proceeds freely through the nebula, the gradual evolution of the orbital radius implies that neither the diffusion coefficient nor the convergence time remain fixed. The task, then, is to revisit the analysis while allowing for their explicit dependence on time.

Strictly speaking, the form of the orbital decay depends on the assumed surface density profile. For $s = 1/2$ and constant aspect ratio $h/r$, the decay is purely exponential. In contrast, for our adopted profile with $s = 3/5$, the solution follows a power law. Nevertheless, within the relevant regime (i.e., where $r$ remains large compared to the disk’s truncation radius and the limit of free migration applies) the resulting decay is well approximated by an exponential. For analytic convenience, we therefore adopt this form and write:
\begin{align}
a = a_0\,\exp\!\bigg[-\frac{t}{\tau_a^0}\,\frac{1+\zeta^2}{1-\zeta^2}\bigg],
\end{align}
where the dependence on $\zeta$ enters because $\tau_a$ characterizes orbital convergence, while the exponential factor reflects the decay timescale of the resonant pair.

Adopting a notation similar to \citet{BatAd2017}, we define the frequency ratio $\bar{\chi} = \Omega_2/\Omega_1 - k/(k-1)$. In terms of this variable, the stochastic differential equation governing the resonant evolution takes the form:
\begin{align}
d\bar{\chi} = \frac{3}{2}\sqrt{2\,\mathcal{D}}\,dw - \frac{3}{2}\frac{\bar{\chi}}{\tau_a}\,dt,
\end{align}
where $w$ denotes a Wiener process. As in earlier work, we take $\bar{\chi}(0)=0$ as the initial condition, and interpret the standard deviation of the resulting distribution as a quantitative measure of stochastic evolution.

In the absence of orbital decay (so that $a=a_0$ and $\mathcal{D}, \tau_a$ are constant), the standard deviation of the distribution function would saturate at \citep{BatAd2017}:
\begin{align}
\delta\bar{\chi}_0 = \sqrt{\frac{3\,\mathcal{D}\,\tau_a}{2}} = \frac{h}{r}\sqrt{\frac{3}{4}\,\frac{\alpha\,\Sigma\,a_0^2\,\chi_a}{m_2\,(1-\zeta)}}.
\label{dchifiducial}
\end{align}
For comparison, the characteristic resonance width is given by \citep{Batygin2015}:
\begin{align}
\Delta\bar{\chi} \approx 5\bigg(\frac{\sqrt{k}\,m_2\,(1+\zeta)}{M}\bigg)^{2/3}.
\label{reswidthchi}
\end{align}
The ratio $\delta\bar{\chi}/\Delta\bar{\chi}\gtrsim1$ therefore provides a convenient analytic criterion for turbulent disruption of resonance.

When orbital decay is included, $\mathcal{D}$ and $\tau_a$ are time-dependent and $\delta\bar{\chi}$ no longer saturates, but continues to evolve with time. Relative to the baseline estimate (\ref{dchifiducial}), the corresponding solution follows the analytic form:
\begin{align}
\frac{\delta\,\bar{\chi}}{\delta\,\bar{\chi}_0} = \frac{3}{625}\sqrt{\frac{6}{35}}\, \bigg(&-390625\,e^{-35\varphi/18}-703125\,e^{-5\varphi/3} -1518750\,e^{-25\varphi/18}-4100625\,e^{-10\varphi/9}-14762250\,e^{-5\varphi/6} \nonumber \\
&-79716150\,e^{-5\varphi/9}-860934420\,e^{-5\varphi/18}+ e^{-\frac{54}{5} e^{5\varphi/18}} 
\times \big( 962125945\,e^{54/5} -9298091736\,\mathrm{Ei} \big(\tfrac{54}{5}\big) \nonumber \\
&+9298091736\,\mathrm{Ei} \big(\tfrac{54}{5} e^{5\varphi/18}\big) \big)\bigg)^{1/2},
\end{align}
where $\varphi = t/\tau_a$ and $\mathrm{Ei} = \int_{-\infty}^x (e^t/t)\,dt$ is the exponential integral. In this case, we take the temporal maximum of $\delta\bar{\chi}$ -- representing the largest departure from the resonant center during migration -- as the relevant quantity for evaluating the disruption threshold. This refinement alters the fiducial estimate only modestly, introducing a correction factor of roughly\footnote{The maximum of this quantity occurs at $\varphi = 0.379$ and evaluates to $\delta\,\bar{\chi}/\delta\,\bar{\chi}_0=0.657$.} $2/3$. The resulting criterion for turbulent disruption thus reads:
\begin{align}
\frac{\delta\,\bar{\chi}}{\Delta\,\bar{\chi}} &\approx \frac{1}{15}\,\frac{h}{r}\,\frac{M}{m_2} \sqrt{\frac{3\,\alpha\,\chi_a}{(1-\zeta)\,(1+\zeta)^{4/3}}} \left(\frac{\Sigma\,r^2}{k\,M}\, \sqrt{\frac{\Sigma\,r^2}{m_2}}\right)^{1/3} \gtrsim 1.
\end{align}
Substituting equations (\ref{horX}) and (\ref{SigmaX}) into the above expression for $(h/r)$ and $\Sigma$ respectively, we obtain equation (\ref{turbulentX}) of the main text.



\begin{thebibliography}


\bibitem[Adams et al.(2008)]{2008ApJ...683.1117A} Adams, F.~C., Laughlin, G., \& Bloch, A.~M.\ 2008, \apj, 683, 2, 1117. doi:10.1086/589986


\bibitem[Armitage(2011)]{2011ARA&A..49..195A} Armitage, P.~J.\ 2011, \araa, 49, 1, 195. doi:10.1146/annurev-astro-081710-102521

\bibitem[Armitage(2019)]{2019SAAS...45....1A} Armitage, P.~J.\ 2019, Saas-Fee Advanced Course, 45, 1. doi:10.1007/978-3-662-58687-7\_1

\bibitem[Ataiee \& Kley(2021)]{2021A&A...648A..69A} Ataiee, S. \& Kley, W.\ 2021, \aap, 648, A69. doi:10.1051/0004-6361/202038772



\bibitem[Balbus \& Hawley(1991)]{1991ApJ...376..214B} Balbus, S.~A. \& Hawley, J.~F.\ 1991, \apj, 376, 214. doi:10.1086/170270


\bibitem[Batygin \& Morbidelli(2013)]{2013AJ....145....1B} Batygin, K. \& Morbidelli, A.\ 2013, \aj, 145, 1, 1. doi:10.1088/0004-6256/145/1/1


\bibitem[Batygin(2015)]{Batygin2015} Batygin, K.\ 2015, \mnras, 451, 3, 2589. doi:10.1093/mnras/stv1063


\bibitem[Batygin et al.(2015)]{2015AJ....149..167B} Batygin, K., Deck, K.~M., \& Holman, M.~J.\ 2015, \aj, 149, 5, 167. doi:10.1088/0004-6256/149/5/167

\bibitem[Batygin \& Adams(2017)]{BatAd2017} Batygin, K. \& Adams, F.~C.\ 2017, \aj, 153, 3, 120. doi:10.3847/1538-3881/153/3/120

\bibitem[Batygin \& Morbidelli(2020)]{2020ApJ...894..143B} Batygin, K. \& Morbidelli, A.\ 2020, \apj, 894, 2, 143. doi:10.3847/1538-4357/ab8937

\bibitem[Batygin \& Morbidelli(2022)]{2022A&A...666A..19B} Batygin, K. \& Morbidelli, A.\ 2022, \aap, 666, A19. doi:10.1051/0004-6361/202243196

\bibitem[Batygin \& Morbidelli(2023)]{2023NatAs...7..330B} Batygin, K. \& Morbidelli, A.\ 2023, Nature Astronomy, 7, 330. doi:10.1038/s41550-022-01850-5


\bibitem[Batygin \& Petit(2023)]{2023ApJ...946L..11B} Batygin, K. \& Petit, A.~C.\ 2023, \apjl, 946, 1, L11. doi:10.3847/2041-8213/acc015


\bibitem[Batygin et al.(2023)]{BatAdBeck23} Batygin, K., Adams, F.~C., \& Becker, J.\ 2023, \apjl, 951, 1, L19. doi:10.3847/2041-8213/acdb5d


\bibitem[Bell \& Lin(1994)]{1994ApJ...427..987B} Bell, K.~R. \& Lin, D.~N.~C.\ 1994, \apj, 427, 987. doi:10.1086/174206

\bibitem[Birnstiel et al.(2010)]{Birnstiel2010} Birnstiel, T., Dullemond, C.~P., \& Brauer, F.\ 2010, \aap, 513, A79. doi:10.1051/0004-6361/200913731

\bibitem[Birnstiel et al.(2011)]{Birnstiel2011} Birnstiel, T., Ormel, C.~W., \& Dullemond, C.~P.\ 2011, \aap, 525, A11. doi:10.1051/0004-6361/201015228


\bibitem[Bitsch et al.(2015)]{2015A&A...575A..28B} Bitsch, B., Johansen, A., Lambrechts, M., et al.\ 2015, \aap, 575, A28. doi:10.1051/0004-6361/201424964

\bibitem[Blandford \& Payne(1982)]{1982MNRAS.199..883B} Blandford, R.~D. \& Payne, D.~G.\ 1982, \mnras, 199, 883. doi:10.1093/mnras/199.4.883


\bibitem[Brunton \& Batygin(2025)]{2025ApJ...991...15B} Brunton, I.~R. \& Batygin, K.\ 2025, \apj, 991, 1, 15. doi:10.3847/1538-4357/adf432



\bibitem[Carter et al.(2012)]{2012Sci...337..556C} Carter, J.~A., Agol, E., Chaplin, W.~J., et al.\ 2012, Science, 337, 6094, 556. doi:10.1126/science.1223269

\bibitem[Charalambous et al.(2018)]{2018MNRAS.477.1414C} Charalambous, C., Mart{\'\i}, J.~G., Beaug{\'e}, C., et al.\ 2018, \mnras, 477, 1, 1414. doi:10.1093/mnras/sty676

\bibitem[Chatterjee \& Tan(2014)]{2014ApJ...780...53C} Chatterjee, S. \& Tan, J.~C.\ 2014, \apj, 780, 1, 53. doi:10.1088/0004-637X/780/1/53


\bibitem[Chiang \& Laughlin(2013)]{2013MNRAS.431.3444C} Chiang, E. \& Laughlin, G.\ 2013, \mnras, 431, 4, 3444. doi:10.1093/mnras/stt424

\bibitem[Choksi \& Chiang(2020)]{2020MNRAS.495.4192C} Choksi, N. \& Chiang, E.\ 2020, \mnras, 495, 4, 4192. doi:10.1093/mnras/staa1421



\bibitem[Dai et al.(2023)]{2023AJ....165...33D} Dai, F., Masuda, K., Beard, C., et al.\ 2023, \aj, 165, 2, 33. doi:10.3847/1538-3881/aca327

\bibitem[Dai et al.(2024)]{2024AJ....168..239D} Dai, F., Goldberg, M., Batygin, K., et al.\ 2024, \aj, 168, 6, 239. doi:10.3847/1538-3881/ad83a6


\bibitem[Deck et al.(2012)]{Deck2012} Deck, K.~M., Holman, M.~J., Agol, E., et al.\ 2012, \apjl, 755, 1, L21. doi:10.1088/2041-8205/755/1/L21

\bibitem[Deck \& Batygin(2015)]{2015ApJ...810..119D} Deck, K.~M. \& Batygin, K.\ 2015, \apj, 810, 2, 119. doi:10.1088/0004-637X/810/2/119

\bibitem[Delisle et al.(2015)]{2015A&A...579A.128D} Delisle, J.-B., Correia, A.~C.~M., \& Laskar, J.\ 2015, \aap, 579, A128. doi:10.1051/0004-6361/201526285

\bibitem[Desch \& Turner(2015)]{2015ApJ...811..156D} Desch, S.~J. \& Turner, N.~J.\ 2015, \apj, 811, 2, 156. doi:10.1088/0004-637X/811/2/156


\bibitem[Dominik \& Dullemond(2008)]{DominikDullemond2008} Dominik, C. \& Dullemond, C.~P.\ 2008, \aap, 491, 3, 663. doi:10.1051/0004-6361:20077493




\bibitem[Elbakyan et al.(2020)]{2020A&A...637A...5E} Elbakyan, V.~G., Johansen, A., Lambrechts, M., et al.\ 2020, \aap, 637, A5. doi:10.1051/0004-6361/201937198

\bibitem[Emsenhuber et al.(2021)]{2021A&A...656A..70E} Emsenhuber, A., Mordasini, C., Burn, R., et al.\ 2021, \aap, 656, A70. doi:10.1051/0004-6361/202038863




\bibitem[Flock et al.(2019)]{2019A&A...630A.147F} Flock, M., Turner, N.~J., Mulders, G.~D., et al.\ 2019, \aap, 630, A147. doi:10.1051/0004-6361/201935806


\bibitem[Friedland(2001)]{2001ApJ...547L..75F} Friedland, L.\ 2001, \apjl, 547, 1, L75. doi:10.1086/318880

\bibitem[Fulton et al.(2017)]{2017AJ....154..109F} Fulton, B.~J., Petigura, E.~A., Howard, A.~W., et al.\ 2017, \aj, 154, 3, 109. doi:10.3847/1538-3881/aa80eb



\bibitem[Goldberg \& Batygin(2021)]{2021AJ....162...16G} Goldberg, M. \& Batygin, K.\ 2021, \aj, 162, 1, 16. doi:10.3847/1538-3881/abfb78


\bibitem[Goldberg \& Batygin(2022)]{2022AJ....163..201G} Goldberg, M. \& Batygin, K.\ 2022, \aj, 163, 5, 201. doi:10.3847/1538-3881/ac5961

\bibitem[Goldreich(1965)]{1965MNRAS.130..159G} Goldreich, P.\ 1965, \mnras, 130, 159. doi:10.1093/mnras/130.3.159

\bibitem[Goldreich \& Schlichting(2014)]{Goldreich2014} Goldreich, P. \& Schlichting, H.~E.\ 2014, \aj, 147, 2, 32. doi:10.1088/0004-6256/147/2/32

\bibitem[Gomes(1995)]{1995Icar..115...47G} Gomes, R.~S.\ 1995, \icarus, 115, 1, 47. doi:10.1006/icar.1995.1077

\bibitem[Go{\'z}dziewski et al.(2016)]{2016MNRAS.455L.104G} Go{\'z}dziewski, K., Migaszewski, C., Panichi, F., et al.\ 2016, \mnras, 455, 1, L104. doi:10.1093/mnrasl/slv156



\bibitem[Hansen \& Murray(2012)]{2012ApJ...751..158H} Hansen, B.~M.~S. \& Murray, N.\ 2012, \apj, 751, 2, 158. doi:10.1088/0004-637X/751/2/158


\bibitem[Hartmann et al.(1998)]{1998ApJ...495..385H} Hartmann, L., Calvet, N., Gullbring, E., et al.\ 1998, \apj, 495, 1, 385. doi:10.1086/305277


\bibitem[Hayashi(1981)]{1981PThPS..70...35H} Hayashi, C.\ 1981, Progress of Theoretical Physics Supplement, 70, 35. doi:10.1143/PTPS.70.35

\bibitem[Henrard \& Lemaitre(1983)]{HenrardLemaitre1983} Henrard, J. \& Lemaitre, A.\ 1983, Celestial Mechanics, 30, 2, 197. doi:10.1007/BF01234306


\bibitem[Howard(2013)]{2013Sci...340..572H} Howard, A.~W.\ 2013, Science, 340, 6132, 572. doi:10.1126/science.1233545

\bibitem[Huang \& Ormel(2022)]{2022MNRAS.511.3814H} Huang, S. \& Ormel, C.~W.\ 2022, \mnras, 511, 3, 3814. doi:10.1093/mnras/stac288


\bibitem[Huang \& Ormel(2023)]{2023MNRAS.522..828H} Huang, S. \& Ormel, C.~W.\ 2023, \mnras, 522, 1, 828. doi:10.1093/mnras/stad1032

\bibitem[H{\"u}hn et al.(2021)]{2021A&A...656A.115H} H{\"u}hn, L.-A., Pichierri, G., Bitsch, B., et al.\ 2021, \aap, 656, A115. doi:10.1051/0004-6361/202142176




\bibitem[Izidoro et al.(2017)]{Izidoro2017} Izidoro, A., Ogihara, M., Raymond, S.~N., et al.\ 2017, \mnras, 470, 2, 1750. doi:10.1093/mnras/stx1232



\bibitem[Jeffreys(1949)]{Jeffreys1949} Jeffreys, H.\ 1949, \textit{Nature}, 163, 262



\bibitem[Kant(1755)]{Kant} Kant, I.\ 1755, \textit{Allgemeine Naturgeschichte und Theorie des Himmels} (Königsberg: Petersen)

\bibitem[Kajtazi et al.(2023)]{2023A&A...669A..44K} Kajtazi, K., Petit, A.~C., \& Johansen, A.\ 2023, \aap, 669, A44. doi:10.1051/0004-6361/202244460

\bibitem[Klahr \& Bodenheimer(2003)]{2003ApJ...582..869K} Klahr, H.~H. \& Bodenheimer, P.\ 2003, \apj, 582, 2, 869. doi:10.1086/344743

\bibitem[Klahr \& Hubbard(2014)]{2014ApJ...788...21K} Klahr, H. \& Hubbard, A.\ 2014, \apj, 788, 1, 21. doi:10.1088/0004-637X/788/1/21


\bibitem[Lambrechts et al.(2019)]{2019A&A...627A..83L} Lambrechts, M., Morbidelli, A., Jacobson, S.~A., et al.\ 2019, \aap, 627, A83. doi:10.1051/0004-6361/201834229


\bibitem[Laplace(1796)]{Laplace}Laplace, P.-S.\ 1796, \textit{Exposition du système du monde} (Paris: Cercle-Social)

\bibitem[Lee \& Peale(2002)]{2002ApJ...567..596L} Lee, M.~H. \& Peale, S.~J.\ 2002, \apj, 567, 1, 596. doi:10.1086/338504

\bibitem[Lee et al.(2014)]{2014ApJ...797...95L} Lee, E.~J., Chiang, E., \& Ormel, C.~W.\ 2014, \apj, 797, 2, 95. doi:10.1088/0004-637X/797/2/95

\bibitem[Lee \& Chiang(2017)]{2017ApJ...842...40L} Lee, E.~J. \& Chiang, E.\ 2017, \apj, 842, 1, 40. doi:10.3847/1538-4357/aa6fb3

\bibitem[Lesur \& Latter(2016)]{2016MNRAS.462.4549L} Lesur, G.~R.~J. \& Latter, H.\ 2016, \mnras, 462, 4, 4549. doi:10.1093/mnras/stw2172


\bibitem[Lithwick \& Wu(2012)]{2012ApJ...756L..11L} Lithwick, Y. \& Wu, Y.\ 2012, \apjl, 756, 1, L11. doi:10.1088/2041-8205/756/1/L11

\bibitem[Lissauer(1993)]{1993ARA&A..31..129L} Lissauer, J.~J.\ 1993, \araa, 31, 129. doi:10.1146/annurev.aa.31.090193.001021

\bibitem[Lopez \& Fortney(2013)]{2013ApJ...776....2L} Lopez, E.~D. \& Fortney, J.~J.\ 2013, \apj, 776, 1, 2. doi:10.1088/0004-637X/776/1/2

\bibitem[Luque et al.(2023)]{2023Natur.623..932L} Luque, R., Osborn, H.~P., Leleu, A., et al.\ 2023, \nat, 623, 7989, 932. doi:10.1038/s41586-023-06692-3

\bibitem[Lynden-Bell \& Pringle(1974)]{1974MNRAS.168..603L} Lynden-Bell, D. \& Pringle, J.~E.\ 1974, \mnras, 168, 603. doi:10.1093/mnras/168.3.603

\bibitem[Lyra \& Umurhan(2019)]{2019PASP..131g2001L} Lyra, W. \& Umurhan, O.~M.\ 2019, \pasp, 131, 1001, 072001. doi:10.1088/1538-3873/aaf5ff




\bibitem[MacDonald et al.(2021)]{2021AJ....162..114M} MacDonald, M.~G., Shakespeare, C.~J., \& Ragozzine, D.\ 2021, \aj, 162, 3, 114. doi:10.3847/1538-3881/ac12d5

\bibitem[Marcus et al.(2015)]{2015ApJ...808...87M} Marcus, P.~S., Pei, S., Jiang, C.-H., et al.\ 2015, \apj, 808, 1, 87. doi:10.1088/0004-637X/808/1/87

\bibitem[Marschall \& Morbidelli(2023)]{2023A&A...677A.136M} Marschall, R. \& Morbidelli, A.\ 2023, \aap, 677, A136. doi:10.1051/0004-6361/202346616


 \bibitem[Masset et al.(2006)]{Masset2006} Masset, F.~S., Morbidelli, A., Crida, A., et al.\ 2006, \apj, 642, 1, 478. doi:10.1086/500967

\bibitem[Message(1966)]{1966IAUS...25..197M} Message, P.~J.\ 1966, The Theory of Orbits in the Solar System and in Stellar Systems, 25, 197. 

\bibitem[Millholland \& Laughlin(2019)]{2019NatAs...3..424M} Millholland, S. \& Laughlin, G.\ 2019, Nature Astronomy, 3, 424. doi:10.1038/s41550-019-0701-7

\bibitem[Millholland et al.(2020)]{2020ApJ...897....7M} Millholland, S., Petigura, E., \& Batygin, K.\ 2020, \apj, 897, 1, 7. doi:10.3847/1538-4357/ab959c

\bibitem[Millholland \& Winn(2021)]{2021ApJ...920L..34M} Millholland, S.~C. \& Winn, J.~N.\ 2021, \apjl, 920, 2, L34. doi:10.3847/2041-8213/ac2c77

\bibitem[Mills et al.(2016)]{2016Natur.533..509M} Mills, S.~M., Fabrycky, D.~C., Migaszewski, C., et al.\ 2016, \nat, 533, 7604, 509. doi:10.1038/nature17445


\bibitem[Morbidelli et al.(2022)]{2022NatAs...6...72M} Morbidelli, A., Bailli{\'e}, K., Batygin, K., et al.\ 2022, Nature Astronomy, 6, 72. doi:10.1038/s41550-021-01517-7


\bibitem[Moulton(1905)]{1905ApJ....22..165M} Moulton, F.~R.\ 1905, \apj, 22, 165. doi:10.1086/141260



\bibitem[Nelson et al.(2013)]{2013MNRAS.435.2610N} Nelson, R.~P., Gressel, O., \& Umurhan, O.~M.\ 2013, \mnras, 435, 3, 2610. doi:10.1093/mnras/stt1475


\bibitem[Nesvorn{\'y} et al.(2025)]{2025AJ....170..180N} Nesvorn{\'y}, D., Morbidelli, A., Bottke, W.~F., et al.\ 2025, \aj, 170, 3, 180. doi:10.3847/1538-3881/adf20a



\bibitem[Ogihara et al.(2024)]{2024ApJ...972..181O} Ogihara, M., Morbidelli, A., \& Kunitomo, M.\ 2024, \apj, 972, 2, 181. doi:10.3847/1538-4357/ad65d5

\bibitem[Okuzumi \& Ormel(2013)]{2013ApJ...771...43O} Okuzumi, S. \& Ormel, C.~W.\ 2013, \apj, 771, 1, 43. doi:10.1088/0004-637X/771/1/43



\bibitem[Paardekooper et al.(2013)]{2013MNRAS.434.3018P} Paardekooper, S.-J., Rein, H., \& Kley, W.\ 2013, \mnras, 434, 4, 3018. doi:10.1093/mnras/stt1224

\bibitem[Papaloizou \& Larwood(2000)]{2000MNRAS.315..823P} Papaloizou, J.~C.~B. \& Larwood, J.~D.\ 2000, \mnras, 315, 4, 823. doi:10.1046/j.1365-8711.2000.03466.x


\bibitem[Peale(1976)]{Peale1976} Peale, S.~J.\ 1976, \araa, 14, 215. doi:10.1146/annurev.aa.14.090176.001243

\bibitem[Peale(1986)]{1986sats.book..159P} Peale, S.~J.\ 1986, IAU Colloquium 77: Some Background about Satellites, 159. 

\bibitem[Petigura et al.(2018)]{2018AJ....155...89P} Petigura, E.~A., Marcy, G.~W., Winn, J.~N., et al.\ 2018, \aj, 155, 2, 89. doi:10.3847/1538-3881/aaa54c

\bibitem[Petigura et al.(2020)]{2020AJ....159....2P} Petigura, E.~A., Livingston, J., Batygin, K., et al.\ 2020, \aj, 159, 1, 2. doi:10.3847/1538-3881/ab5220

\bibitem[Petit et al.(2020)]{2020MNRAS.496.3101P} Petit, A.~C., Petigura, E.~A., Davies, M.~B., et al.\ 2020, \mnras, 496, 3, 3101. doi:10.1093/mnras/staa1736

\bibitem[Pichierri et al.(2018)]{2018CeMDA.130...54P} Pichierri, G., Morbidelli, A., \& Crida, A.\ 2018, Celestial Mechanics and Dynamical Astronomy, 130, 8, 54. doi:10.1007/s10569-018-9848-2


\bibitem[Pichierri et al.(2024)]{2024NatAs...8.1408P} Pichierri, G., Morbidelli, A., Batygin, K., et al.\ 2024, Nature Astronomy, 8, 1408. doi:10.1038/s41550-024-02342-4


\bibitem[Press et al.(1992)]{1992nrfa.book.....P} Press, W.~H., Teukolsky, S.~A., Vetterling, W.~T., Flannery, B.~P.\ 1992.\ Numerical recipes in FORTRAN. The art of scientific computing.\ (Cambridge: University Press)


\bibitem[Quillen et al.(2013)]{Quillen2013} Quillen, A.~C., Bodman, E., \& Moore, A.\ 2013, \mnras, 435, 3, 2256. doi:10.1093/mnras/stt1442



\bibitem[Raymond et al.(2018)]{2018MNRAS.479L..81R} Raymond, S.~N., Boulet, T., Izidoro, A., et al.\ 2018, \mnras, 479, 1, L81. doi:10.1093/mnrasl/sly100


\bibitem[Rein et al.(2010)]{2010A&A...510A...4R} Rein, H., Papaloizou, J.~C.~B., \& Kley, W.\ 2010, \aap, 510, A4. doi:10.1051/0004-6361/200913208


\bibitem[Rimlinger \& Hamilton(2021)]{2021MNRAS.501.4255R} Rimlinger, T. \& Hamilton, D.\ 2021, \mnras, 501, 3, 4255. doi:10.1093/mnras/staa3933

\bibitem[Rosotti et al.(2017)]{2017MNRAS.468.1631R} Rosotti, G.~P., Clarke, C.~J., Manara, C.~F., et al.\ 2017, \mnras, 468, 2, 1631. doi:10.1093/mnras/stx595




\bibitem[Safronov(1969)]{Safronov1969} Safronov, V.~S.\ 1969, \textit{Evoliutsiia doplanetnogo oblaka} (Moscow: Nauka)

\bibitem[Sengupta et al.(2019)]{2019ApJ...874...26S} Sengupta, D., Dodson-Robinson, S.~E., Hasegawa, Y., et al.\ 2019, \apj, 874, 1, 26. doi:10.3847/1538-4357/aafc36


\bibitem[Shakura \& Sunyaev(1973)]{1973A&A....24..337S} Shakura, N.~I. \& Sunyaev, R.~A.\ 1973, \aap, 24, 337. 


\bibitem[Sinclair(1970)]{1970MNRAS.148..325S} Sinclair, A.~T.\ 1970, \mnras, 148, 325. doi:10.1093/mnras/148.3.325


\bibitem[Stevenson(1982)]{1982P&SS...30..755S} Stevenson, D.~J.\ 1982, \planss, 30, 8, 755. doi:10.1016/0032-0633(82)90108-8


\bibitem[Swedenborg(1734)]{Swedenborg} Swedenborg, E.\ 1734, \textit{Opera Philosophica et Mineralia} (Dresden \& Leipzig: Frolich)



\bibitem[Tabone et al.(2022)]{2022MNRAS.512.2290T} Tabone, B., Rosotti, G.~P., Cridland, A.~J., et al.\ 2022, \mnras, 512, 2, 2290. doi:10.1093/mnras/stab3442


\bibitem[Tanaka et al.(2002)]{2002ApJ...565.1257T} Tanaka, H., Takeuchi, T., \& Ward, W.~R.\ 2002, \apj, 565, 2, 1257. doi:10.1086/324713

\bibitem[Tanaka \& Ward(2004)]{2004ApJ...602..388T} Tanaka, H. \& Ward, W.~R.\ 2004, \apj, 602, 1, 388. doi:10.1086/380992


\bibitem[Ter Haar(1967)]{1967ARA&A...5..267T} Ter Haar, D.\ 1967, \araa, 5, 267. doi:10.1146/annurev.aa.05.090167.001411

\bibitem[Teyssandier et al.(2015)]{2015MNRAS.452.1743T} Teyssandier, J., Owen, J.~E., Adams, F.~C., et al.\ 2015, \mnras, 452, 2, 1743. doi:10.1093/mnras/stv1386





\bibitem[Ward(1997)]{1997Icar..126..261W} Ward, W.~R.\ 1997, \icarus, 126, 2, 261. doi:10.1006/icar.1996.5647


\bibitem[Weiss et al.(2018)]{2018AJ....155...48W} Weiss, L.~M., Marcy, G.~W., Petigura, E.~A., et al.\ 2018, \aj, 155, 1, 48. doi:10.3847/1538-3881/aa9ff6

\bibitem[Weiss et al.(2024)]{2024ApJS..270....8W} Weiss, L.~M., Isaacson, H., Howard, A.~W., et al.\ 2024, \apjs, 270, 1, 8. doi:10.3847/1538-4365/ad0cab


\bibitem[Weizs{\"a}cker(1943)]{1943ZA.....22..319W} Weizs{\"a}cker, C.~F.~V.\ 1943, \zap, 22, 319. 

\bibitem[Williams \& Cieza(2011)]{2011ARA&A..49...67W} Williams, J.~P. \& Cieza, L.~A.\ 2011, \araa, 49, 1, 67. doi:10.1146/annurev-astro-081710-102548

\bibitem[Woitke et al.(2016)]{2016A&A...586A.103W} Woitke, P., Min, M., Pinte, C., et al.\ 2016, \aap, 586, A103. doi:10.1051/0004-6361/201526538





\bibitem[Yap \& Batygin(2024)]{2024Icar..41716085Y} Yap, T.~E. \& Batygin, K.\ 2024, \icarus, 417, 116085. doi:10.1016/j.icarus.2024.116085

\bibitem[Yap et al.(2025)]{2025PSJ.....6....2Y} Yap, T.~E., Batygin, K., \& Tissot, F.~L.~H.\ 2025, PSJ, 6, 1, 2. doi:10.3847/PSJ/ad92fa


\bibitem[Youdin(2011)]{2011ApJ...742...38Y} Youdin, A.~N.\ 2011, \apj, 742, 1, 38. doi:10.1088/0004-637X/742/1/38


\bibitem[Zeng et al.(2016)]{2016ApJ...819..127Z} Zeng, L., Sasselov, D.~D., \& Jacobsen, S.~B.\ 2016, \apj, 819, 2, 127. doi:10.3847/0004-637X/819/2/127

\bibitem[Zhang \& Bellan(2022)]{2022ApJ...930..167Z} Zhang, Y. \& Bellan, P.~M.\ 2022, \apj, 930, 2, 167. doi:10.3847/1538-4357/ac62d5


\end{thebibliography}

\bibliographystyle{aasjournal}

\end{document}